\documentclass{aastex631}

\shorttitle{AASTeX v6.3.1 Sample article}
\shortauthors{Peña-Asensio et al.}
\graphicspath{{./}{figures/}}

\begin{document}

\title{Orbital characterization of superbolides observed from space: dynamical association with near-Earth objects, meteoroid streams and identification of hyperbolic projectiles}

\author[0000-0002-7257-2150]{Eloy Peña-Asensio}
\affiliation{Departament de Química, Universitat Autònoma de Barcelona 08193 Bellaterra, Catalonia, Spain}
\affiliation{Institut de Ciències de l’Espai (ICE, CSIC) Campus UAB, C/ de Can Magrans s/n, 08193 Cerdanyola del Vallès, Catalonia, Spain}

\author[0000-0001-8417-702X]{Josep M. Trigo-Rodríguez}
\affiliation{Institut de Ciències de l’Espai (ICE, CSIC) Campus UAB, C/ de Can Magrans s/n, 
08193 Cerdanyola del Vallès, Catalonia, Spain}
\affiliation{Institut d’Estudis Espacials de Catalunya (IEEC) 08034 Barcelona, Catalonia, Spain}

\author[0000-0002-9637-4554]{Albert Rimola}
\affiliation{Departament de Química, Universitat Autònoma de Barcelona 08193 Bellaterra, Catalonia, Spain}



\begin{abstract}

There is an unceasing incoming flux of extraterrestrial materials reaching the Earth's atmosphere. Some of these objects produce luminous columns when they ablate during the hypersonic encounter with air molecules. A few fireballs occur each year bright enough to be detected from space. The source of these events is still a matter of debate, but it is generally accepted that they are of sporadic origin. We studied the NASA-JPL Center for NEOs Studies (CNEOS) fireball database to infer the dynamic origin of large bolides produced by meter-sized projectiles that impacted our planet. These likely meteorite-dropping events were recorded by the US Government satellite sensors. We estimated the false-positive rate and analyzed the time evolution of multiple orbit dissimilarity criteria concerning potential associations with near-Earth objects and meteoroid streams. We found that at least 16\% of the large bolides could be associated with meteoroid streams, about 4\% are likely associated with near-Earth asteroids, and 4\% may be linked to near-Earth comets. This implies that a significant fraction of meter-sized impactors producing large bolides may have an asteroidal or cometary origin. In addition, we found at least three bolides having hyperbolic orbits with high tensile strength values. Meter-sized meteoroids of interstellar origin could be more common than previously thought, representing about 1\% of the flux of large bolides. The inferred bulk physical properties suggest that the interstellar medium could bias these projectiles towards high strength rocks with the ability to survive prolonged exposure to the harsh interstellar space conditions.

\end{abstract}



\section{Introduction} \label{sec:intro}

The Earth is bombarded annually by dozens of meter-sized objects from space \citep{Brown2002Natur.420..294B}. These bodies impact the atmosphere at hypervelocity radiating enormous amounts of energy due to the ablation produced when colliding with the air molecules \citep{Revelle1979JATP...41..453R}. The largest bodies can survive this process and reach the Earth's surface generating dangerous scenarios as exemplified by Chelyabinsk \citep{Brown2013Natur.503..238B}, being also the source of meteorite falls \citep{Ceplecha1998SSRv...84..327C, Gritsevich2012CosRe..50...56G, Sansom2019ApJ...885..115S} and more rarely of high strength projectiles capable to excavate impact craters \citep{Tancredi2009M&PS...44.1967T}. The associated risk depends on different factors, mainly the size and mass of the object, but also the geometry of incidence and the velocity with respect to the Earth \citep{Baldwin1971JGR....76.4653B}. Knowing the nature and origin of these impactors is of vital importance to prevent a localized catastrophe, as well as studying the hazardous range of sizes associated with small asteroids and comets, which remain largely undiscovered to date \citep{Mainzer2011ApJ...743..156M, Granvik2018Icar..312..181G}. In this sense, the analysis of their dynamic origin from all the available data, including space sensors, allows us to delve into the sources and the physical mechanisms producing dangerous projectiles.

Fortunately, events capable of globally devastating our planet occur on scales of millions of years \citep{Morrison2019Icar..327....1M, Trigo2022asteroid}. The largest well-documented impact was the Tunguska event in 1908, a projectile of around 60 m that released as much energy as hundreds of atomic bombs \citep{Morrison2019Icar..327....1M, Robertson2019Icar..327...36R}. More recently there was the well-known Chelyabinsk event, a 19-meter asteroid that went unnoticed by all observation instruments and exploded over a Russian city, injuring thousands of people \citep{Brown2013Natur.503..238B, Boro2016A&A...585A..90B, Kartashova2018P&SS..160..107K}.

However, meter-sized Earth impacting objects are more common that we could think from our experience. On average, about 30 asteroids of a few meters in diameter are detected each year colliding with our planet, which are the least observed objects in the Solar System due to their size and low albedo for the dominant chondritic population. Hypervelocity causes that these rocky, rocky-metal or metallic bodies generate tens of kilometer long luminous columns of ionized gas and dust during the atmospheric entry, producing the so-called superbolides when they are brighter than magnitude -17 \citep{Ceplecha1999md98.conf...37C, Koschny2017JIMO...45...91K}. These events release so much energy (around 1 kilotons of TNT in average) that their light curves can be detected by satellite sensors. The detections from space are useful and complementary to the data obtained from the ground by fireball networks \citep{Trigo2004EM&P...95..553T, Jenniskens2011Icar..216...40J, Howie2017ExA....43..237H, Colas2020A&A...644A..53C, Eloy2021MNRAS.504.4829P}.

The origin of these projectiles is still under discussion since there are multiple factors involved in the orbital evolution as well as the diversity of possible sources. Dynamically associating bodies can be highly challenging even more for rare events such large bolides. In this regard, the purpose of the present work is to investigate the orbits of meter-sized hazardous projectiles that impacted our atmosphere, and to explore the connection they may have with cometary and asteroidal meteoroid streams and near-Earth Objects (NEOs). Taking into account relevant previous work on the dynamics of these rocks \citep{Pauls2005MandPS...40.1241P, Fu2005Icar..178..434F, Schunova2012Icar..220.1050S}, our main goal is to find if there is a significant population of transient meter-sized projectiles directly associated with asteroids or comets belonging to the NEO population based on space-borne observation. The mere existence of such associations is noteworthy because, so far, we have understood that most meteorite-dropping events are sporadic in origin. CNEOS data properly analyzed, having into account uncertainties associated with the detections and generating a synthetic population to compare each event against the random background, will demonstrate that a substantial part of these meter-sized hazardous populations could be younger and might be produced in shorter timescales, being a consequence of the physical processes that experience NEOs in near-Earth space. In addition, the data analysis has allowed us to identify a small population of likely interstellar meteoroids, at least three confidently characterized as exhibiting unequivocally hyperbolic orbital parameters.

\section{Database and methodology} \label{sec:datamet}

The NASA-JPL Center for NEOs Studies (CNEOS) has been monitoring atmospheric flares produced by bright bolides since 1988 with space sensors. These records are collected by the Nuclear Test Ban Treaty monitoring satellites, also known as US Government (USG) sensors, a classified military system \citep{Tagliaferri1994hdtc.conf..199T}.

The data provided by CNEOS are generally in good agreement with independent ground-based records in location and time. However, for some cases, important differences in velocity have been reported \citep{Boro2017P&SS..143..147B, Granvik2018Icar..311..271G, Devillepoix2019MNRAS.483.5166D}. This may be because space sensors are most likely to capture events in the largest subtended atmospheric volume, i.e., away from the most effective detection area of the sensors. Previously, these data (59 events up to 2015) have already been partially used to analyze size-frequency of meter-scaled objects \citep{Brown2002Natur.420..294B} and to examine their physical properties and orbital parameters \citep{Brown2016Icar..266...96B}.

\citet{Brown2016Icar..266...96B} did not find any clear association, even though 17\% of the studied events exhibited typical Jupiter Family Comet (JFC) orbits. To test meteoroid stream linkages, they used the Drummond criterion \citep{Drummond1981Icar...45..545D} for orbit dissimilarity with a cut-off of 0.05 assuming an error in CNEOS velocity around 0.1-0.2 km/s. They pointed to the Taurid shower and its sub-components as the only significant associated shower in the dataset studied. However, \citet{Devillepoix2019MNRAS.483.5166D} stated that this velocity error may not be realistic and reported some variations above 20\% having deviations in the radiant of up to 90º when checked with ground-based measurements.

These complications in the data analysis greatly hinders finding associations with parent bodies using the CNEOS database. First, because of the lack of uncertainty information, and second, because the observed velocity vector is a crucial factor in calculating meteoroid orbits, which in some cases exhibit large errors \citet{Devillepoix2019MNRAS.483.5166D}. For this reason, following a normal distribution, we assume a standard deviation in latitude and longitude of 0.1 degrees, in heights of 0.5 km and 20\% for the components of the velocity vector. From this, we apply a Monte Carlo simulation by generating normal distributions on each parameter, randomly producing 10,000 clones for each event. Using our \textit{3D-FireTOC} software \citep{Eloy2021MNRAS.504.4829P, Eloy2021Falcon9}, we compute each of the clone orbital elements, which are compared with the entire database of The International Astronomical Union (IAU) \citep{Jopek2014me13.conf..353J, Jopek2017P&SS..143....3J, Jenniskens2020P&SS..18204821J}, using the established meteor shower (EMS), the working meteor shower (WMS), the near-Earth comet  (NEC) list, and the near-Earth asteroid (NEA) dataset, each of which is composed of 110, 615, 121, and 25,798 unique entries respectively.

Since we assume a distribution of uncorrelated errors following a normal distribution, we find that some of the clones have spurious orbits, as would be expected since especially the uncertainties of the velocity vector components are most likely correlated. To identify these anomalous data and filter with the same criteria, we apply a statistical approach typically used to detect outliers in a univariate dataset that approximately follows a normal distribution: the Generalized Extreme Studentized Deviate (GESD) method \citep{rosner1983percentage}. We clean our clone population with an upper limit of 1,000 outliers at a 5\% significance level, re-generating new clones to always maintain the initial population size. In addition, as the computed orbital elements exhibit a skewed normal distributions, we use the median and the median absolute deviation indicators instead of the mean and standard deviation, which are less robust for asymmetric distributions.

Different criteria have been developed to measure the similarity of orbits based on semi-quantitative formulas. These functions were created to measure how an orbit can evolve into another due to perturbations. They are, therefore, approximations, and the criteria thresholds, to some extent, arbitrary. Depending on the inclination of the orbit or the dispersion of the orbital elements of a complex, the threshold values may vary. However, it should only be considered reducing the cut-off values if the number of low inclination associations requires it.

The first orbital dissimilarity function we use is the one established by \citet{Southworth1963SCoA....7..261S}:

\begin{equation}
D_{S H}^{2}=\left(e_{B}-e_{A}\right)^{2}+\left(q_{B}-q_{A}\right)^{2}+\left(2 \sin \frac{I_{AB}}{2}\right)^{2} \\
+\left(\frac{e_{B}+e_{A}}{2}\right)^{2}\left(2 \sin \frac{\pi_{B A}}{2}\right)^{2},
	\label{eq:D_SH}
\end{equation}

where $e$ is the eccentricity, $q$ is the perihelion distance, $I_{AB}$ is the angle made by the orbit's planes and $\pi_{BA}$ the difference between longitudes of perihelia measured from the intersection of both orbits.

By adding weights in the first parts of the equation \ref{eq:D_SH}, \citet{Drummond1981Icar...45..545D} modified the formula reaching the following expression:

\begin{equation}
D_{D}^{2}=\left(\frac{e_{B}-e_{A}}{e_{B}+e_{A}}\right)^{2}+\left(\frac{q_{B}-q_{A}}{q_{B}+q_{A}}\right)^{2}+\left(\frac{I_{AB}}{\pi}\right)^{2} \\
+\left(\frac{e_{B}+e_{A}}{2}\right)^{2}\left(\frac{\theta_{B A}}{\pi}\right)^{2},
	\label{eq:D_D}
\end{equation}

where $\theta_{BA}$ is the orbit angle between the lines of apsides.

As a combination of equations \ref{eq:D_SH} and \ref{eq:D_D}, \citet{Jopek1993Icar..106..603J} suggested a new criterion:

\begin{equation}
D_{H}^{2}=\left(e_{B}-e_{A}\right)^{2}+\left(\frac{q_{B}-q_{A}}{q_{B}+q_{A}}\right)^{2}+\left(2 \sin \frac{I_{AB}}{2}\right)^{2} \\
+\left(\frac{e_{B}+e_{A}}{2}\right)^{2}\left(2 \sin \frac{\pi_{B A}}{2}\right)^{2}.
	\label{eq:D_H}
\end{equation}

Finally, we also consider the criterion proposed by \citet{Jenniskens2008Icar..194...13J}, which obtained some invariants derived from the observed orbital elements of a meteoroid stream or a parent body:

\begin{equation}
D_{J}^{2}=\left( \frac{C 1_{1}-C 1_{2}} { 0.13}\right)^{2}+\left( \frac{C 2_{1}-C 2_{2}}{0.06}\right)^{2}+\left(\frac{C 3_{1}-C 3_{2}}{14.2^{\circ}}\right)^{2}.
	\label{eq:D_J}
\end{equation}

In equation \ref{eq:D_J}, $C1$, $C2$ and $C3$ can be computed as:

\begin{eqnarray}
    C1 & = &   (1-e^2) \cos^2(i), \nonumber\\
    C2 & = &  e^2(0.4-\sin^2(i)\sin^2(\omega)),\\
    C3 & = &  \omega + \Omega, \nonumber
\end{eqnarray}

where $\omega$ is the argument of perihelion and $\Omega$ is the longitude of the ascending node.

Table \ref{tab:criteria} lists the four different criteria we use and their corresponding thresholds (in their broadest options due to the large uncertainties of the data).

\begin{table*}
	\centering
	\caption{Threshold criterion typically used for orbit dissimilarity and their references.}
	\label{tab:criteria}
    \begin{tabular}{lcccc}
    	\hline
    
        Criterion & Ref. Criterion                 & Threshold & Ref. Threshold                \\
        \hline
        $D_{SH}$     & \citep{Southworth1963SCoA....7..261S}             & 0.3       & \citep{Porub2006CoSka..36..103P}      \\
        $D_J$        & \citep{Jenniskens2008Icar..194...13J}             & 1.5       & \citep{Jenniskens2008Icar..194...13J}            \\
        $D_D$        & \citep{Drummond1981Icar...45..545D}               & 0.18      & \citep{Moorhead2019JIMO...47..134M}               \\
        $D_H$        & \citep{Jopek1993Icar..106..603J}                  & 0.35      & \citep{Jopek1997AandA...320..631J} \\
        \hline

    \end{tabular}
\end{table*}

As a first approximation, we count as possibly associated every clone that at least meets one of these criteria. To check the robustness of the associations, considering that the most influential parameter in the orbit is the velocity modulus, we separate the clones in different ranges according to their observed velocity. In this way, we can check how strong the association is as the pre-atmospheric velocity varies. This gives a first idea of the probability that an event may be associated with a meteor shower or a parent body even though the uncertainty is unknown. If an event is mostly associated with an object in all velocity variation ranges, the linkage likelihood would be high. However, if an event is associated in some ranges, not in others, or only appeared associated exclusively when the velocity modulus varied greatly, the likelihood that it belongs to such a complex or originates from a certain object would be lower.

When working with large datasets, the probability of two orbits being similar at any time by coincidence is considerably high \citep{Southworth1963SCoA....7..261S, Wiegert2004EM&P...95...19W, Porub2004EM&P...95..697P}. Therefore, we need a quantitative metric to indicate the rate of false positives for every event with each of the compared datasets, so we can be convinced that the rate of association is well above that expected by chance. We generate a synthetic population of 1,000 orbits for each observed CNEOS event following \citet{Pauls2005MandPS...40.1241P} by generating a semi-major axis, eccentricity and inclination distribution from the individual orbital elements computed, drawing from this in a Monte Carlo sense and assigning random ascending node and argument of perihelion values with the constrains that the generated orbit intercepts the Earth. For each synthetic impactor the same 10,000 cloning process is performed and then comparison with the reference meteoroid stream, NEA and NEC populations performed to find the false positive rate. This provides a clear grounding as to whether or not the statistical association rates found are significant given the thresholds adopted. Associations that do not exceed the randomly expected mean are directly discarded to provide only the most likely candidates. However, it is not a sufficient condition that the association exceeds the expected at random. We perform null-hypothesis significance testing and use the p-value as a measure equivalent to the false-positive rate, i.e., we calculate the probability of the associations not being random.

To impose the condition that the orbit impacts the Earth, we set random longitude of node and set an argument of perihelion that fulfills the condition:

\begin{equation}
\frac{1}{e} \left( \frac{a(1-e^2)}{r_{min}} \right) < \cos\omega < \frac{1}{e} \left( \frac{a(1-e^2)}{r_{max}} \right),
	\label{eq:imp_cond}
\end{equation}

where $r_{min}$ and $r_{max}$ represent the inner radius of 0.983 AU and the outer radius of 1.017 AU of the torus defined by the Earth's orbit and which must necessarily contain the ascending or descending node of any orbit that eventually intersects our planet.

We then count all the clones associated with each meteoroid stream or parent body that meet at least one criterion and are associated with more than 75\% of the clones, and we sort the associations by frequency. Following \citet{Porub2004EM&P...95..697P}, to check that the orbits are not just similar at the time of impact by coincidence and to verify the consistency of these links over time, we numerically integrate every CNEOS event and parent body candidates 10,000 years backward in time, representing parent meteoroid stream candidates with 18 particles uniformly distributed throughout each stream orbit. Since the original parent body may have fragments scattered along the whole orbit, it is important to integrate particles for each stream spread in true anomaly, as bodies belonging to the same complex may diverge due to close encountering with planets \citep{Dmitriev2015P&SS..117..223D}. In this way, we can check how the dissimilarity criteria evolve for different positions of the meteor shower orbit. In this regard, we use the IAS 15 integrator implemented in the \textit{REBOUND} package \citep{Rein2012A&A...537A.128R}, including the Sun and all the planets in the Solar System. We track the evolution of each dissimilarity criterion over time considering a linkage as robust when it remains below the cut-off for 5,000 years. 

Typically, the bulk density of meteoroids entering the atmosphere is estimated by calculating the tensile strength ($s$) at the time of flare. This is usually done by following \citet{bronshten1981physics} approximation ($s\approx\rho_{atm}V^2 $), which is the pressure in the shock layer that has caused the disruption of the object at a certain height and velocity. However, a more sophisticated fragmentation model is needed since the analysis of superbolides and the comparison with recovered meteorites prove that large meteoroids or small asteroids undergo atmospheric disruption processes at dynamical pressure lower than their mechanical strength \citep{Popova2011M&PS...46.1525P}. \citet{Foschini2001A&A...365..612F} considered the shock wave turbulence interaction that locally increases the dynamic pressure exerted on the meteoroid and approximated the tensile strength as:

\begin{equation}
s \approx (1+\alpha)\rho_{atm}\kappa V^2,
	\label{eq:tensile}
\end{equation}

where $\alpha$ is the degree of ionization, $\kappa$ is an amplification factor of the kinetic energy ($2 \leq \kappa \leq 6$ for monoatomic gases and plasma), and $\rho_{atm}$ and $V^2$ are the atmospheric density and the velocity respectively at the time of disruption. However, we observed that this approximation offers values far in excess of those expected for tensile strength given the values provided by CNEOS. Alternatively, we use the following fragmentation model:

\begin{equation}
s \approx \frac{(\gamma-1)(1+\alpha)}{2\gamma}\rho_{atm}V^2,
	\label{eq:tensile2}
\end{equation}

where $\alpha=1$ since the fluid at the stagnation point is highly ionized, and $\gamma=1.7$ is the ratio of specific heats \citep{Kadono1996JGR...10126097K, Foschini1999A&A...342L...1F, Farinella1998Icar..132..378F}. The value of the atmospheric density is derived from the data of U.S. standard atmosphere \citep{atmosphere1976national}. 

From this, we classify the body as cometary if $s<10^5\,Pa$, carbonaceous if  $10^5\,Pa<s<10^6\,Pa$, rocky if $10^6\,Pa<s<10^7\,Pa$, and rocky-iron if its tensile strength is greater \citep{Chyba1993Natur.361...40C}. This will allow us to compare the densities with the Tisserand parameter with respect to Jupiter ($T_j$) and to compute the size of the object assuming that the energy recorded by the USG sensors is equal to the kinetic energy as follows:

\begin{equation}
D = 2 \sqrt[3]{\frac{3T_E}{2\pi\rho v^2}}.
	\label{eq:D}
\end{equation}

With this approach, we intend both to account for the unknown uncertainty of the CNEOS detections and to encompass phenomena that eventually differentiate the orbit of an object from its complex and that may have generated, precisely, an exceptional event (in terms of mass and energy) apparently unrelated to its initial swarm. By using this methodology, we will obtain the most likely candidates in case an event is truly associated with a  or a NEO.

Calculations have been performed by using the CSUC (Consorci de Serveis Universitaris de Catalunya) supercomputing infrastructures, employing 2,500 hours of computation on 100 Intel Xeon Platinum 8168 central processing unit at 2.7 GHz.

\section{RESULTS}

As of March 1, 2022, the CNEOS public database counted 887 events starting from 1988. However, only 255 contained sufficient data to reconstruct their orbit (i.e, date, longitude, latitude, height, and observed velocity vector).

\begin{figure}[ht!]
	\plotone{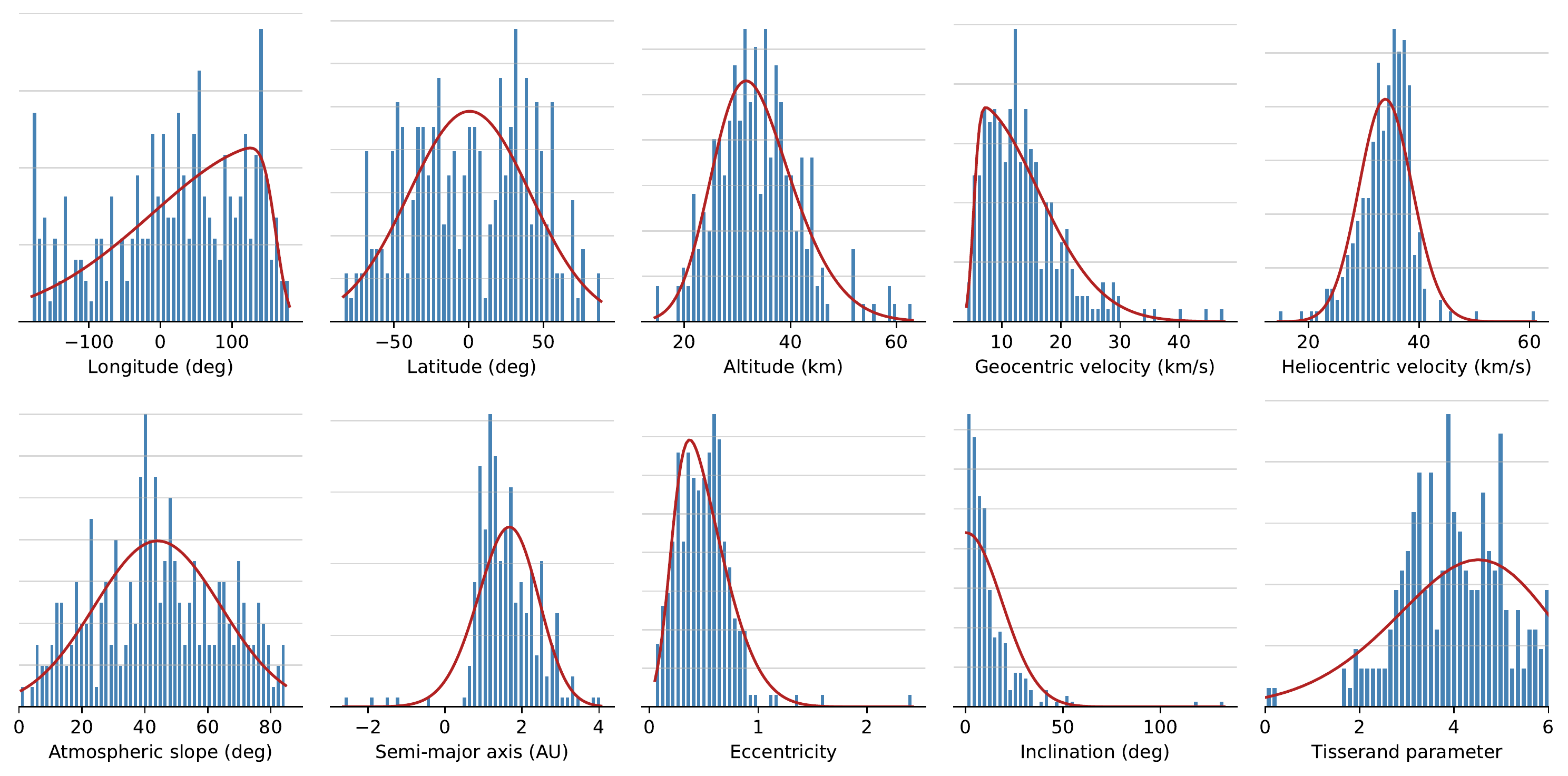}
    \caption{Histogram of the principal parameters and the calculated orbits of the 255 CNEOS studied events in this work. The y-axis is arbitrarily scaled.}
    \label{fig:hist_orbits}
\end{figure}

The fireballs recorded have a random distribution around the Earth and seem to come from all parts of the sky, that is, the radiants are distributed throughout the celestial sphere. The fireballs appear to have a random impact angle with respect to the local horizon and a random azimuth. Figure \ref{fig:hist_orbits} shows the principal parameters and the calculated orbits of the studied events.

\begin{figure}[ht!]
	\plotone{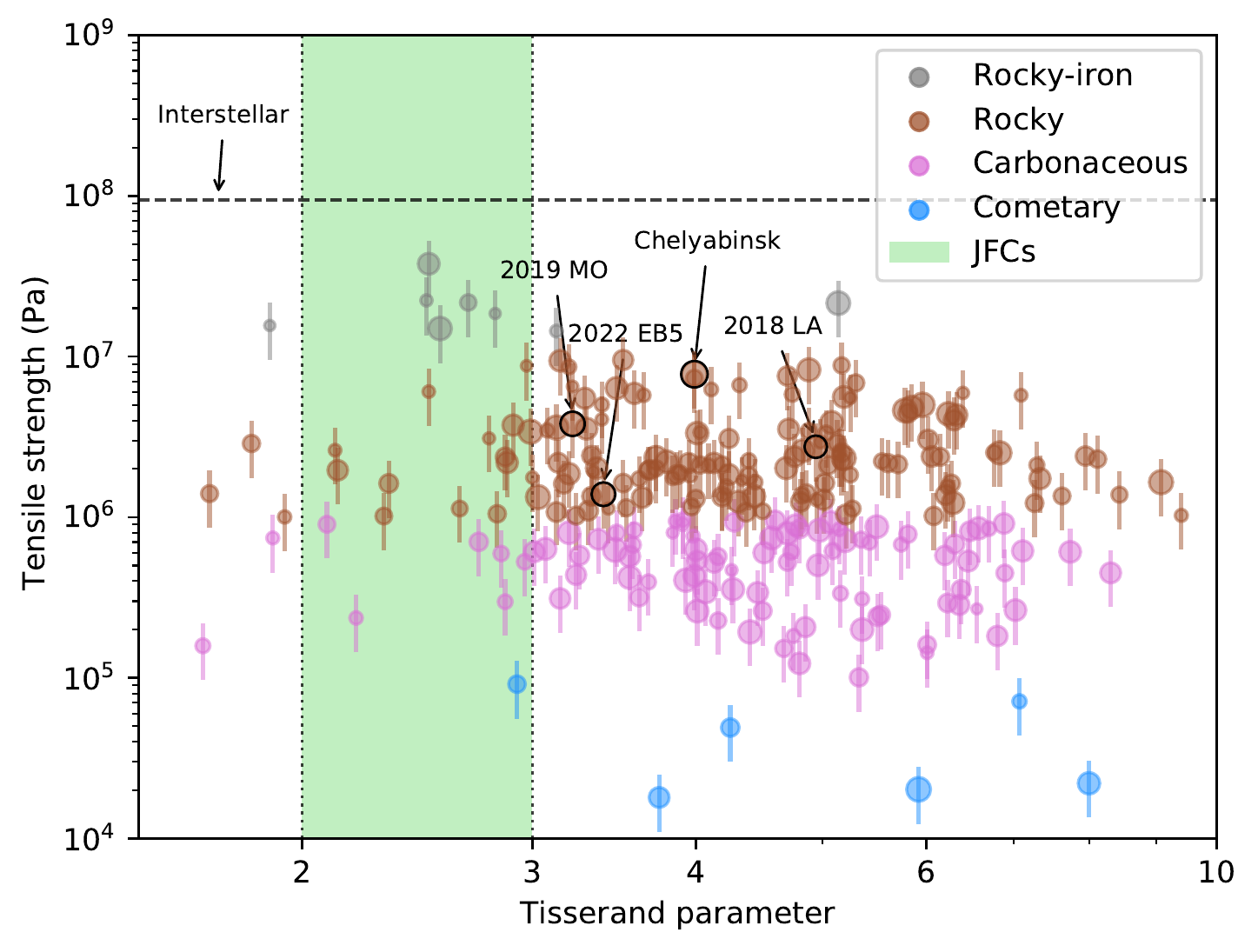}
    \caption{Tensile strength as a function of the Tisserand parameter for the CNEOS events. The radii of the circles are arbitrarily scaled with the diameter of the object. Impacts corresponding to the asteroid 2022 EB5, 2019 MO, 2018 LA, the Chelyabinsk event and the first interstellar meteoroid are annotated.}
    \label{fig:tensile}
\end{figure}

As a first indication of the possible origin of each event, we calculate the tensile strength and classify them according to the above mentioned criteria in Section \ref{sec:datamet}. We find that 35.7\% of the dynamic strengths correspond to carbonaceous bodies, 2.4\% exhibit bulk densities typical of cometary meteoroids, while 61.9\% appears to be rocky or rocky-iron bodies. On the other hand, 9.8\% of the events belong to the JFCs, while 85.5\% of them have values of the Tisserand parameter characteristic of asteroid-like orbits ($T_j>3$). In Figure \ref{fig:tensile} is depicted the tensile strength as a function of the Tisserand parameter with its respective classification. The size of the circles refers to the diameter of the body which, as can be seen, does not seem to reveal any correlation. Like expected, the Chelyabinsk superbolide produced by a LL5 ordinary chondrite falls between the rocky and rocky-iron dominion \citep{Galimov2013SoSyR..47..255G,Zaytsev2021EM&P..125....2Z}.

We chose the best associations since some of the events have two or more candidates from the same dataset. We impose as a minimum criterion that the association appears in at least in three-fourths of the clone population. Table \ref{tab:asso_summary} shows the number of event associations to meteor showers or NEOs according to the number of criteria they comply with and differentiating between the established and the working list (5 events have multiple candidates from cross-datasets).

\begin{deluxetable*}{lccccc}
\tablecaption{Best meteoroid stream, NEA and NEC associations counts that appear in more than 75\% of the generated random clones, exceed the expected random mean and fulfill criteria for at least 5,000 years.\label{tab:asso_summary}}
\tablewidth{0pt}
\tablehead{
\colhead{Assoc.} & \colhead{1 crit.}  & \colhead{2 crit.} &
\colhead{3 crit.} & \colhead{4 crit.} & \colhead{All crit.} 
}
\startdata
    EMS   & 1       & 2       & 1       & 13      & 17        \\
    WMS   & 2       & 3       & 4       & 15      & 24        \\
    NEA   & 0       & 2       & 1       & 8       & 11        \\
    NEC   & 5       & 1       & 2       & 3       & 11        \\
    Total & 8       & 8       & 8      & 39      & 63       \\
\enddata
\end{deluxetable*}

The number of repetitions of the association, i.e., the number of clones associated with an object, can be understood as the probability that the event actually belongs to that specific meteoroid stream (or parent body) assuming a 20\% uncertainty on the observed velocity vector provided by the CNEOS database. Tables \ref{tab:EMS}, \ref{tab:WMS}, \ref{tab:NEA} and \ref{tab:NEC} group the events by i) established, ii) working meteor shower, and iii) NEA and NEC associations, respectively, together with the number of criteria they meet, the percentage of associations in the clones both in total and by velocity variation range, and the probability of not being random.

\begin{deluxetable*}{lccccccc}
\tablecaption{Best established meteor shower associations by event.\label{tab:EMS}}
\tablewidth{0pt}
\tablehead{
\colhead{Established meteor shower} & \colhead{Event}  & \colhead{Nº crit.} &
\colhead{$R_{Total}$} & \colhead{$R_{0-5\%}$} & \colhead{$R_{5\%-10\%}$} & \colhead{$R_{10\%-20\%}$}  & \colhead{p-value}
}
\startdata
Corvids & 2005-04-19 07:37:47 & 4 & 96.56\% & 98.57\% & 98.85\% & 96.69\% & 93.81\% \\
 & 2019-09-12 12:49:48 & 4 & 91.77\% & 99.88\% & 99.96\% & 95.45\% & 71.85\% \\
 & 2019-06-22 21:25:48 & 4 & 93.91\% & 99.97\% & 99.79\% & 88.08\% & 65.11\% \\
 & 2019-07-23 20:42:58 & 4 & 85.43\% & 99.96\% & 99.92\% & 70.75\% & 60.54\% \\
 & 2013-04-21 06:23:12 & 4 & 80.09\% & 99.89\% & 93.80\% & 53.02\% & 56.02\% \\
 & 2018-06-26 17:51:53 & 4 & 77.99\% & 84.45\% & 82.83\% & 65.52\% & 50.00\% \\
 & 2006-09-02 04:26:15 & 3 & 87.29\% & 98.85\% & 95.72\% & 62.53\% & 67.50\% \\
 & 2018-09-25 14:10:33 & 2 & 85.55\% & 99.87\% & 99.91\% & 94.68\% & 89.17\% \\
 & 2019-05-19 14:47:03 & 2 & 82.40\% & 92.44\% & 87.73\% & 69.32\% & 67.42\% \\
Daytime kappa Aquariids & 2005-06-03 08:15:41 & 4 & 97.84\% & 99.50\% & 98.44\% & 98.11\% & 83.81\% \\
h Virginids & 2015-04-21 01:42:51 & 4 & 93.02\% & 94.48\% & 95.32\% & 91.87\% & 75.51\% \\
 & 2006-10-14 18:10:49 & 4 & 98.77\% & 99.88\% & 99.82\% & 99.56\% & 53.24\% \\
tau Herculids & 2014-06-28 02:40:07 & 4 & 77.87\% & 99.16\% & 89.28\% & 49.44\% & 63.92\% \\
Andromedids & 2020-12-29 20:32:22 & 4 & 87.16\% & 99.57\% & 94.81\% & 62.86\% & 63.30\% \\
Phoenicids & 2006-12-09 06:31:12 & 4 & 76.51\% & 99.49\% & 98.28\% & 58.30\% & 59.71\% \\
alpha Virginids & 2017-02-18 19:48:29 & 4 & 92.32\% & 99.77\% & 99.71\% & 96.90\% & 53.65\% \\
alpha Capricornids & 2005-12-03 12:45:49 & 1 & 78.80\% & 93.52\% & 92.27\% & 53.96\% & 59.49\% \\
\enddata
\end{deluxetable*}

\begin{deluxetable*}{lccccccc}
\tablecaption{Best working meteor shower associations by event.\label{tab:WMS}}
\tablewidth{0pt}
\tablehead{
\colhead{Working meteor shower} & \colhead{Event}  & \colhead{Nº crit.} &
\colhead{$R_{Total}$} & \colhead{$R_{0-5\%}$} & \colhead{$R_{5\%-10\%}$} & \colhead{$R_{10\%-20\%}$}  & \colhead{p-value}
}
\startdata
iota Cygnids & 2008-11-18 09:41:51 & 4 & 82.07\% & 92.01\% & 88.74\% & 75.46\% & 92.95\% \\
Daytime delta Scorpiids & 2014-11-26 23:16:51 & 4 & 96.97\% & 99.06\% & 98.55\% & 96.11\% & 85.78\% \\
 & 2008-06-27 02:01:23 & 4 & 84.88\% & 88.49\% & 85.97\% & 84.74\% & 51.70\% \\
alpha Geminids & 2006-01-10 23:25:28 & 4 & 96.29\% & 96.14\% & 96.57\% & 96.81\% & 78.91\% \\
 & 2013-12-23 08:30:57 & 4 & 98.77\% & 99.97\% & 99.06\% & 98.79\% & 70.87\% \\
 & 2019-09-14 12:39:34 & 4 & 98.91\% & 99.63\% & 99.10\% & 99.41\% & 64.37\% \\
Southern omega Scorpiids & 2007-09-22 17:57:12 & 4 & 91.18\% & 98.94\% & 98.73\% & 81.14\% & 71.63\% \\
 & 2004-10-07 13:14:43 & 1 & 83.55\% & 98.54\% & 95.71\% & 65.85\% & 50.00\% \\
gamma Taurids & 2017-05-24 07:03:03 & 4 & 95.21\% & 96.43\% & 95.86\% & 96.40\% & 70.49\% \\
 & 2014-11-04 20:13:30 & 4 & 95.82\% & 100\% & 98.75\% & 88.92\% & 55.92\% \\
 & 2021-04-02 15:52:58 & 4 & 96.79\% & 99.84\% & 98.86\% & 90.05\% & 54.51\% \\
 & 2018-10-05 00:27:04 & 3 & 92.48\% & 99.14\% & 97.72\% & 94.74\% & 55.26\% \\
 & 2017-07-13 09:30:36 & 2 & 99.07\% & 99.89\% & 98.05\% & 99.02\% & 68.73\% \\
pi Leonids & 2008-11-21 00:26:44 & 4 & 87.42\% & 99.54\% & 95.86\% & 83.42\% & 63.44\% \\
Southern October delta Arietids & 2017-06-23 20:21:55 & 4 & 80.40\% & 82.15\% & 79.77\% & 77.16\% & 60.97\% \\
beta Cancrids & 2014-05-16 20:06:28 & 4 & 87.32\% & 96.42\% & 92.7\% & 80.99\% & 60.36\% \\
gamma Triangulids & 2020-10-26 15:09:10 & 4 & 82.99\% & 96.46\% & 94.58\% & 70.74\% & 56.4\% \\
 & 2015-04-08 04:06:31 & 3 & 92.36\% & 99.64\% & 98.93\% & 94.58\% & 50.00\% \\
October gamma Cetids & 2013-12-08 03:10:09 & 4 & 97.09\% & 100\% & 100\% & 100\% & 54.41\% \\
 & 2013-04-30 08:40:38 & 2 & 84.30\% & 100\% & 100\% & 69.27\% & 54.34\% \\
May lambda Draconids & 2006-06-07 00:06:28 & 3 & 98.04\% & 99.64\% & 99.53\% & 99.20\% & 75.32\% \\
Daytime theta Aurigids & 2020-07-12 07:50:32 & 3 & 82.58\% & 83.80\% & 84.02\% & 81.24\% & 73.96\% \\
Northern chi Orionids & 2017-12-15 13:14:37 & 2 & 84.22\% & 98.36\% & 94.73\% & 79.13\% & 72.89\% \\
gamma Delphinids & 2022-02-17 03:53:24 & 1 & 75.77\% & 94.35\% & 92.25\% & 79.96\% & 81.37\% \\
\enddata
\end{deluxetable*}

\begin{deluxetable*}{lccccccc}
\tablecaption{Best near-Earth asteroid associations by event. $^*$2022-03-11 21:22:46  event satisfies the condition of $D_D$ for 3,400 years and $D_SH$, $D_J$ and $D_H$ for more than 2000 years.  \label{tab:NEA}}
\tablewidth{0pt}
\tablehead{
\colhead{Near-Earth asteroid} & \colhead{Event}  & \colhead{Nº crit.} &
\colhead{$R_{Total}$} & \colhead{$R_{0-5\%}$} & \colhead{$R_{5\%-10\%}$} & \colhead{$R_{10\%-20\%}$}  & \colhead{p-value}
}
\startdata
2021 AU4 & 2020-07-12 07:50:32 & 4 & 89.69\% & 92.97\% & 92.48\% & 89.37\% & 99.99\% \\
2005 CV69 & 2021-03-05 13:50:01 & 4 & 85.26\% & 94.80\% & 92.57\% & 85.97\% & 99.99\% \\
2016 WN8 & 2021-11-28 18:06:50 & 4 & 94.57\% & 97.17\% & 97.51\% & 97.66\% & 93.43\% \\
2015 AO43 & 2019-02-01 18:17:10 & 4 & 82.50\% & 98.79\% & 95.37\% & 68.71\% & 80.16\% \\
2020 DF4 & 2020-03-04 20:25:59 & 4 & 93.51\% & 100\% & 100\% & 99.97\% & 69.03\% \\
1996 JG & 2014-11-26 23:16:51 & 4 & 94.31\% & 97.46\% & 97.21\% & 93.32\% & 67.25\% \\
2014 XO7 & 2010-12-25 23:24:00 & 4 & 98.07\% & 100\% & 100\% & 99.97\% & 59.13\% \\
2012 KN11 & 2021-11-17 15:53:21 & 4 & 83.01\% & 92.04\% & 89.77\% & 79.05\% & 50.74\% \\
2020 KF6 & 2012-05-15 11:04:17 & 3 & 92.16\% & 100\% & 100\% & 99.92\% & 54.64\% \\
2018 BJ5 & 2012-07-25 07:48:20 & 2 & 92.03\% & 100\% & 100\% & 99.32\% & 74.51\% \\
2008 ON13 & 2017-12-15 13:14:37 & 2 & 91.59\% & 99.50\% & 99.06\% & 92.42\% & 50.00\% \\
2022 EB5$^*$ & 2022-03-11 21:22:46 & 0 & 80.44\% & 98.46\% & 87.03\% & 56.80\% & 50.00\% \\
\enddata
\end{deluxetable*}

\begin{deluxetable*}{lccccccc}
\tablecaption{Best near-Earth comet associations by event. \label{tab:NEC}}
\tablewidth{0pt}
\tablehead{
\colhead{Near-Earth comet} & \colhead{Event}  & \colhead{Nº crit.} &
\colhead{$R_{Total}$} & \colhead{$R_{0-5\%}$} & \colhead{$R_{5\%-10\%}$} & \colhead{$R_{10\%-20\%}$}  & \colhead{p-value}
}
\startdata
300P/Catalina & 2019-09-12 12:49:48 & 4 & 85.33\% & 99.88\% & 99.96\% & 85.55\% & 65.23\% \\
 & 2015-07-19 07:06:26 & 1 & 82.50\% & 91.57\% & 88.83\% & 76.03\% & 88.90\% \\
79P/du Toit-Hartley & 2010-02-28 22:24:50 & 4 & 94.51\% & 99.61\% & 98.49\% & 86.34\% & 59.64\% \\
C/1905 F1 & 2020-03-04 20:25:59 & 4 & 99.60\% & 100\% & 100\% & 100\% & 99.99\% \\
 & 2019-03-19 02:06:39 & 3 & 98.07\% & 99.57\% & 99.20\% & 99.07\% & 99.99\% \\
 & 2003-09-27 12:59:02 & 2 & 83.68\% & 97.94\% & 95.50\% & 86.25\% & 99.96\% \\
P/2003 T12 & 2018-01-15 02:18:38 & 3 & 82.26\% & 91.10\% & 90.01\% & 74.71\% & 92.69\% \\
 & 2008-12-24 15:51:58 & 1 & 78.58\% & 89.32\% & 85.16\% & 71.15\% & 99.28\% \\
182P/LONEOS & 2003-11-10 13:54:06 & 1 & 83.70\% & 92.81\% & 90.13\% & 75.90\% & 64.07\% \\
 & 2015-04-08 04:06:31 & 1 & 89.96\% & 98.50\% & 97.98\% & 87.46\% & 77.53\% \\
34D/Gale & 2007-06-11 09:47:05 & 1 & 75.40\% & 99.44\% & 99.34\% & 80.31\% & 52.09\% \\
\enddata
\end{deluxetable*}

\begin{figure}[ht!]
	\plotone{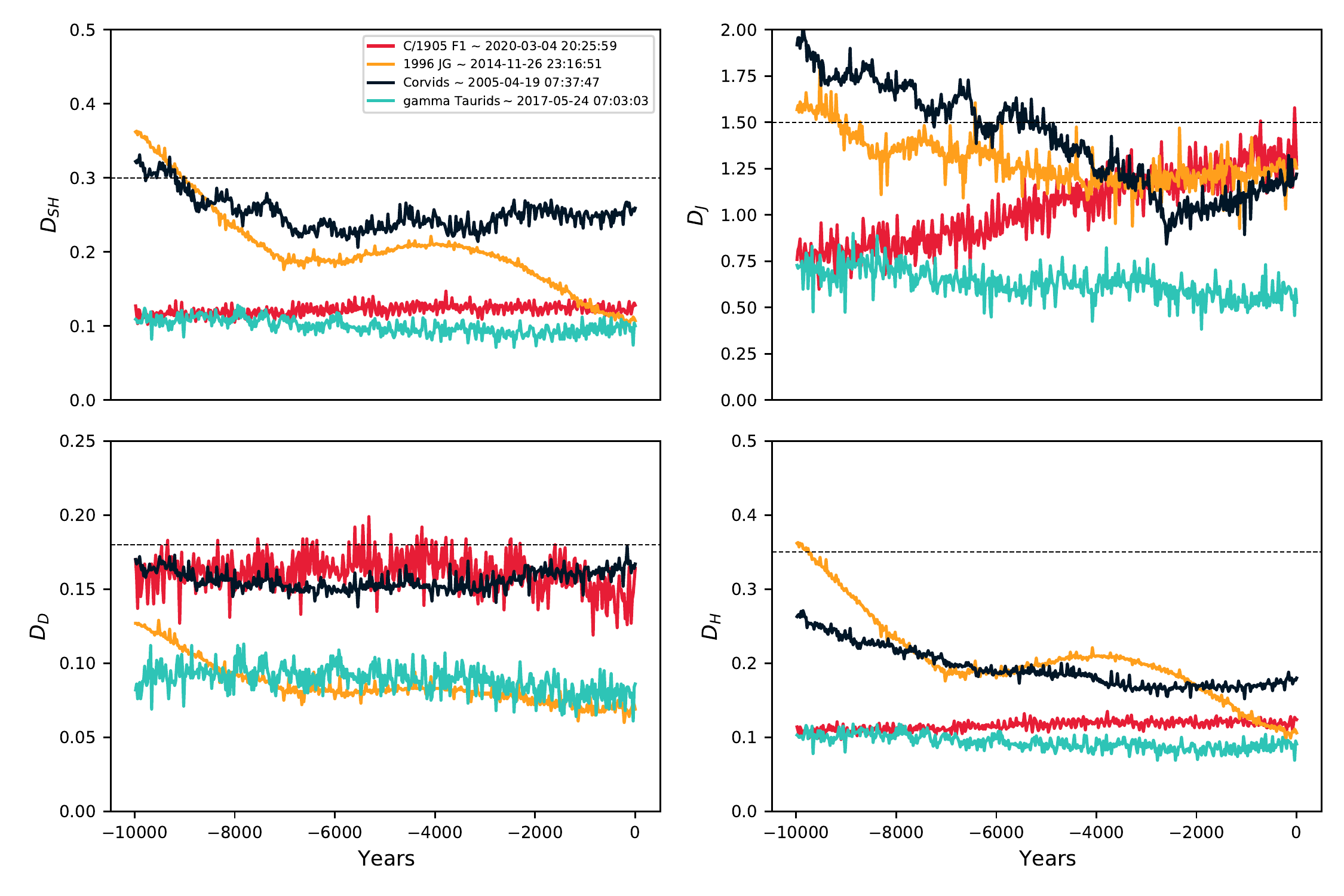}
    \caption{Dissimilarity criteria of 2020-03-04 20:25:59, 2014-11-26 23:16:51, 2005-04-19 07:37:47 and 2017-05-24 07:03:03 events in relation to their respective associations over a backward integration of 10,000 years. A dashed horizontal line shows the cut-off for each criterion.}
    \label{fig:propagation}
\end{figure}

As a sample, in Figure \ref{fig:propagation} we display the evolution of the dissimilarity criteria during the orbital integration for 10,000 years of four events that appear associated with a meteor shower fulfilling the established criteria. As can be seen, some orbits become more similar as time progresses backwards during the first 5,000 years. For convenience, Table \ref{tab:appendix} summarizes the time evolution of the $D_{SH}$ and $D_{D}$ criteria over 10,000 years backwards of all events with possible association.

Out of the 255 events studied, 41 appear to be associated with a unique meteor shower, 11 have an unique NEA association and 11 may be linked to an unique NEC, while 5 events present multiple candidates. This means that almost 23\% of the large fireballs recorded by USG sensors have a likely cometary or asteroidal origin, which according to \citet{Devillepoix2019MNRAS.483.5166D} could occur in more than 60\% of the cases. We find especially relevant the  number of associations with NECs since they appear in around 4\% of the events, equivalent to NEAs (4\%). If the associations were not distinguishable from the background, it would be expected that NEA associations would produce a huge number of candidates since the asteroid dataset used is much larger than the others.

In relation to the concordance between the dissimilarity criteria, Table \ref{tab:crit_relation} shows the percentage of coincidence between the different criteria, as well as the percentage of associations for each one in relation to the total number of associations evaluated.

\begin{deluxetable*}{cccccc}
\tablecaption{Coincidences of associations between the dissimilarity criteria and percentage of associations corresponding to the total of potential associations evaluated.\label{tab:crit_relation}}
\tablewidth{0pt}
\tablehead{
\colhead{} & \colhead{$D_{SH}$} & \colhead{$D_J$}  & \colhead{$D_D$} &
\colhead{$D_H$} & \colhead{Assoc.}  
}
\startdata
    $D_{SH}$   & -       & 69.0\%      & 71.3\%   & 90.2\% & 19.4\%      \\
    $D_J$   & 69.0\%     & -           & 73.6\%   & 68.4\% & 20.0\%     \\
    $D_D$   & 71.3\%     & 73.6\%      & -        & 77.6\% & 20.0\%           \\
    $D_H$   & 90.2\%     & 68.4\%    & 77.6\%   & -      & 20.4\%         \\
\enddata

\end{deluxetable*}

\section{Discussion}

We consider as extremely remarkable the Corvid complex as a source of meter-sized projectiles (linked to 2004 HW asteroid), the Tau Herculids complex (linked to Comet 73P/Schwassman-Wachmann 3) and the h Virginids meteor showers. We also found an association of a bolide with the kappa Aquarids that was previously identified as the source of large bolides \citep{TrigoKappa2007MNRAS.382.1933T}.

Associations with Corvids or h Virginids represent more than half of the total. This overrepresentation of orbits associated with these meteoroid streams is similar to that found in \citet{Dumitru2017A&A...607A...5D}. This is probably due to the fact that they are swarms with very low inclination orbits, and the CNEOS events present an inclination distribution with a tendency to coplanar orbits, as depicted in Figure \ref{fig:hist_orbits}. It is remarkable the dynamical association of comets 300P/Catalina, C/1905 F1 (Giacobini), P/2003 T12 (SOHO) and 182P/LONEOS with multiple large bolides strongly satisfying several criteria. 

\begin{figure}[ht!]
	\plotone{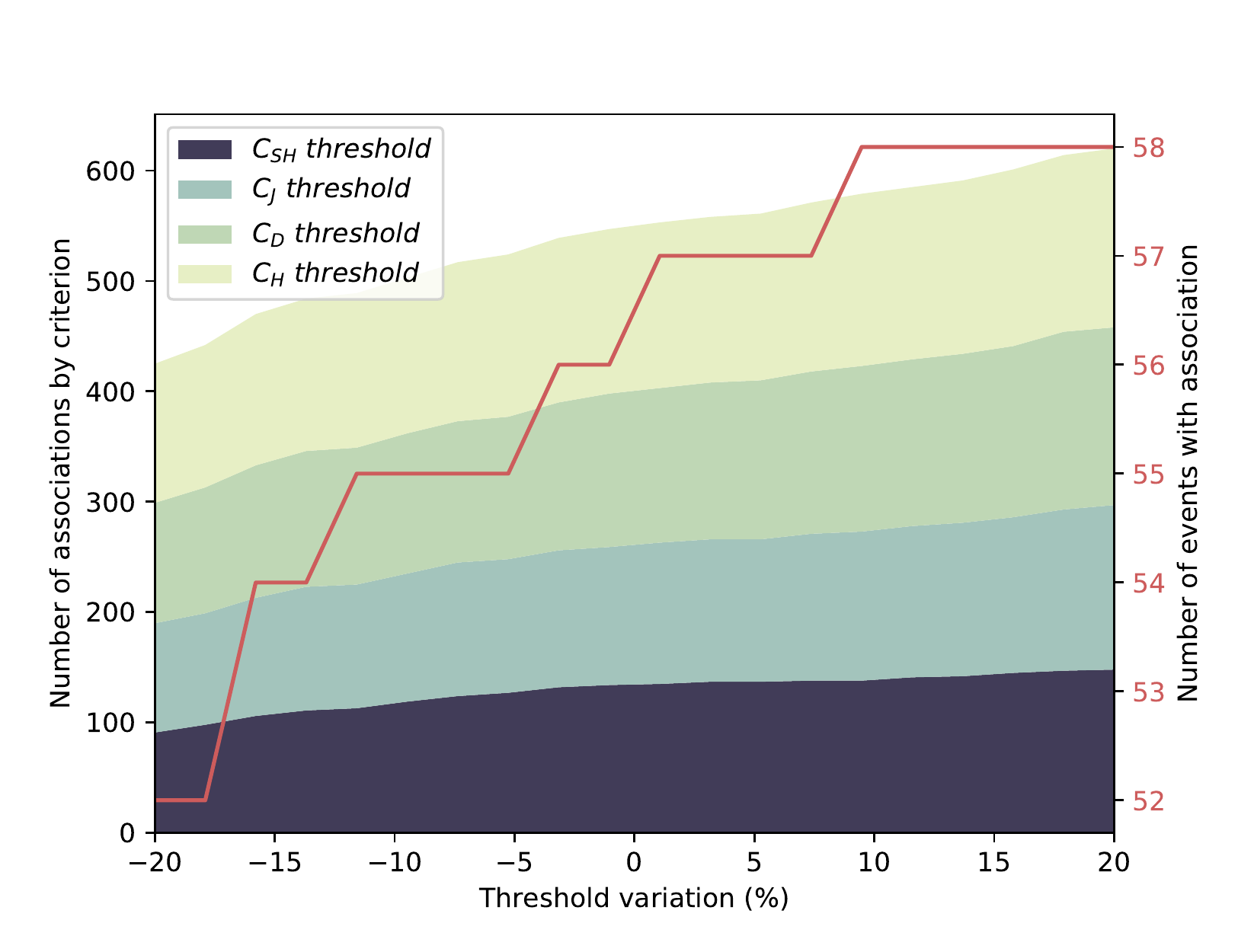}
    \caption{Variation of associations by criterion and events with association when modifying the threshold of the criteria by $\pm$25\%. The number of associations by criterion are presented stacked, however, they do have multiples overlapping.}
    \label{fig:crit_variation}
\end{figure}

Regarding the associations with different velocity deviations, we observed that in most of the obtained cases they are considerably robust, except for a few cases for velocity variations larger than 10\% in modulus. This is because the associations that only appeared in a specific range of velocities were not strong enough to be statistically significant (appearing in more than 75\% of the total clones), thus demonstrating the reliability of the proposed method. To check the robustness of the chosen cut-off values, in Figure \ref{fig:crit_variation} is depicted how the number of associations by criterion and the number of events with associations vary when the thresholds are modified by $\pm$25\%. As can be seen, even with the lowest cut-off for each criterion, the number of associations is still significant: around 20\% of the CNEOS fireball events would have a cometary or asteroidal origin. 

We also note that some events have multiple potential associations, in some cases because the meteoroid streams share parent bodies (such as Andromedids and December Phi Cassiopeiids), while in other cases simply because of the similarity of orbits.

We note that the method applied in this work to calculate the tensile strength possibly results in excessively high values. For example, we obtain 8$\pm$3 MPa for the Chelyabinsk event, while its bulk strength is estimated to be 1 MPa \citep{Boro2013Natur.503..235B}. We can observe that the dynamic pressure computed for the first fulguration at 45 km height is 0.7 MPa, and as it penetrated the atmosphere it grew up to 18 MPa at 22 km, which is in good agreement with our results since the height given by CNEOS of the radiated energy peak is 23.3 km. Therefore, this fact could lead to an overestimation of the tensile strength of almost one order of magnitude.

Concerning the reliability of the dissimilarity criteria, good agreement can be observed between the $D_{J}$ and $D_D$ criteria, which have a reasonable match and a similar percentage of associations, see Table \ref{tab:crit_relation}. Both are below 25\% of associations, which is the estimated likelihood of chance association with random background NEAs \citep{Wiegert2004EM&P...95...19W}. This fact is especially remarkable since both criteria were defined with different theoretical approaches. For subsequent studies, it would be of interest to analyze comparatively the evolution of more dissimilarity criteria, particularly incorporating others like the one proposed by G. Valsecchi that is partially invariant to secular perturbations \citep{Valsecchi1999esra}. The $D_H$ for the chosen threshold values is the one that appears as the most permissive criterion.

It is worth pointing out that some associations do not coincide in date with the expected time of activity for the corresponding meteor shower. This is especially noticeable in meteoroid streams with inclinations close to the ecliptic, such as gamma Taurids. Different works have analyzed the associations between NEOs and meteorite-dropping bolides \citep{Halliday1987Icar...69..550H, Trigo2007MNRAS.382.1933T, Boro2015aste.book..257B}. Recent evidence also come from some meteorite recovered with an accurate orbital determination as e.g. Annama which could be dynamically associated with 2014 UR116 (2008 XB) \citep{Trigo2015MNRAS.449.2119T}, or Chelyabinsk with asteroid (86039) 1999 NC43 \citep{Boro2013Natur.503..235B}. In fact, fragments from these parent bodies result from impacts, but also can be generated by tidal force disruptions as some are rubble-piles and fast rotators \citep{Trigo2007MNRAS.382.1933T, Trigo2009P&SS...57..243T, Chapman2010Natur.463..305C, Trigo2014acm..conf..533T}. Fragmentation processes can produce swarms of meter-sized rocks that can suffer large orbit perturbations over hundreds of years (10$^4$ - 10$^5$ years) \citep{Pauls2005MandPS...40.1241P}. Those complexes whose orbits are highly inclined may be less affected by planetary perturbations and have longer lifetimes \citep{Jones2008EM&P..102...35J, TrigoInclination10.1111}. As a result of the disruption, the different fragments can undergo significant divergences between their orbits, which are only accentuated with the passage of time \citep{Bottke2002Icar..156..399B}. In addition, other non-gravitational effects can alter the orbits being relevant for the dynamic association of objects, for instance, radiation pressure, Yarkovsky and YORP effects, Poynting–Robertson effect or Lorentz force, however, these are only expected to be significant for very small objects over long time scales \citep{Broz2006PhDT.......281B}.

Catastrophic disruptions generally eject fragments at escape velocity, which is much lower than the orbital velocity \citep{Jenniskens1998EP&S...50..555J, Bottke2005Icar..179...63B}. Therefore, many fragments remain in similar orbits producing sources of potential projectiles \citep{Williams2004JIMO...32...11W, Porub2004EM&P...95..697P, Jenniskens2007pimo.conf...56J}. 

The decoherence time scale in meteoroid streams can be so pronounced as to be hardly indistinguishable from a chance association \citep{Pauls2005MandPS...40.1241P}. This can be explained by several mechanisms: (1) the streams may intercept the Earth several times, producing different activity periods due to the joint action of planetary perturbation and collisions with micrometeoroids, which can produce an orbital scattering \citep{Murray1982Icar...49..125M, Babadzhanov1987PAICz..67..141B, TrigoDust2005ApJ...621.1146T}; (2) non-gravitational orbital scattering caused by the Yarkovsky orbital drift derived from solar radiation \citep{Farinella1998Icar..132..378F}; (3) weakly bound agglomerates, such as rubble-piles asteroids or fragile comets, can suffer tidal distortion and disruption if they become fast rotators \citep{Richardson1998Icar..134...47R, Schunova2014Icar..238..156S}; and (4) the effect of the sublimation pressure produced by the cometary nucleus, which may cause meteoroid stream members to undergo mass segregation just as they form. Mass segregation may also result from the spiral approaches of a comet to the Sun producing the release of fragments at different times and velocities \citep{Hughes1981MNRAS.195..625H, Williams1991ASIB..272..225W};

Several bodies from cometary reservoirs beyond Neptune are strongly perturbed by Jupiter or Saturn as they cross their orbits, eventually decaying to Earth-like orbits \citep{Hahn1990Natur.348..132H, Asher1993MNRAS.263..179A}. JFCs are originally trans-Neptunian objects (TNO) and Centaurs that have evolved into the inner planetary region by a progressive decay \citep{Duncan1995AJ....110.3073D, Levison1997Icar..127...13L, Emel2004MNRAS.350..161E, Jewitt2008tnoc.book....1J}. Some of these comets can eventually cross the Earth's orbit \citep{Ipatov2004NYASA1017...46I}. Although many more asteroidal objects impact the atmosphere than JFCs, encounters with the latter can produce much more energetic bolides. It has even been estimated that some of the superbolides produced by meter-sized meteoroids may not necessarily be fragile and could be associated with dormant comets or Damocloids \citep[see e.g.][]{TrigoDormant2017ASSP...46...11T}.

Among the 255 events analyzed, we identify 5 hyperbolic orbits, 3 of which maintain this characteristic in the whole uncertainty range. Table \ref{tab:interstellar} shows these large fireballs with presumably unbound heliocentric orbits and their respective tensile strength, right ascension and declination of the geocentric radiant, heliocentric velocity, semi-major axis, eccentricity and inclination. To confirm that these bodies do indeed come from outside our Solar System, it would be necessary to perform a dynamic analysis over time with trustworthy uncertainty assumptions. \citet{Siraj2019arXiv190407224S} found that 2014-01-08 17:05:34 event had a velocity away from the Local Standard of Rest velocity under large uncertainty variations, announcing the first fireball detection of interstellar origin. Although our results are in good agreement with their computed orbital elements, we notice a discrepancy in the estimation of the radiant position. In addition, 2017-03-09 04:16:37 and 2009-04-10 18:42:45 events also appear to have Solar System unbound orbits within the margins of error. Corroborating these metric-sized interstellar objects impacting our atmosphere would indicate that about a 1\% of these events may be produced by bodies from beyond our heliosphere. As suggested by the high tensile strengths obtained (in the known range of values for high strength chondrites and iron meteorites: $10^7$ to $10^8\,Pa$), and in agreement with \citep{Siraj_2022}, it is perhaps more unlikely that submetric-sized meteoroids survive long stays in interstellar space due to the effect of thermal stress and cosmic radiation over millions of years in comparison with large objects (see Figure \ref{fig:tensile}). The presence of refractory aggregates in the interstellar medium could progressively erode smaller fragile objects biasing cometary meteoroids, which, in turn, could be more common due to their eccentric orbits \citep{TrigoJurgen202210.1093/mnras/stab2827}. The identified CNEOS bolides were produced by roughly meter-sized rocks, but other two larger bodies have been recently identified: 1I/'Oumuamua or 2I/Borisov \citep{Meech2017Natur.552..378M, Guzik2020NatAs...4...53G}. The great difference in the observation frequency of these interstellar fireballs ($\sim$3/255) versus interstellar objects ($\sim$2/1000000) could simply be due to the lower expected number of large bodies compared to the smaller ones. It is also likely that a significant amount of interstellar meteors remains hidden under the uncertainties of ground-based observation networks \citep{Hajdukova2019msme.book..235H, Hajdukova2020P&SS..19004965H}.

\begin{deluxetable*}{lccccccc}
\tablecaption{CNEOS events with hyperbolic orbits and their respective tensile strength, right ascension and declination of the geocentric radiant, heliocentric velocity, semi-major axis, eccentricity and inclination.\label{tab:interstellar}}
\tablewidth{0pt}
\tablehead{
\colhead{Event} & \colhead{$s\, (MPa)$} & \colhead{$RA_{geo}\,(^{\circ})$} & \colhead{$DEC_{geo}\,(^{\circ})$} & \colhead{$V_{h}\, (km/s)$} & \colhead{$a\, (AU)$}  & \colhead{$e$} &
\colhead{ $i\,(^{\circ})$}  
}
\startdata
2021-05-06 05:54:27 & 11.8 $\pm$ 3.4 & 62.6 $\pm$ 5.8 & 12.2 $\pm$ 3.2 & 44.1 $\pm$ 2.5 & -2.64 $\pm$ 3.10 & 1.15 $\pm$ 0.18 & 6.05 $\pm$ 2.56  \\
2017-03-09 04:16:37 & 75.5 $\pm$ 14.9 & 170.7 $\pm$ 6.3 & 34.1 $\pm$ 6.8 & 50.1 $\pm$ 3.8 & -1.22 $\pm$ 0.60 & 1.57 $\pm$ 0.30 & 24.03 $\pm$ 5.57  \\
2015-02-17 13:19:50 & 4.0 $\pm$ 1.5 & 339.3 $\pm$ 9.1 & -9.6 $\pm$ 2.7 & 44.0 $\pm$ 5.1 & -1.45 $\pm$ 4.94 & 1.10 $\pm$ 0.32 & 1.12 $\pm$ 0.99 \\
2014-01-08 17:05:34 & 222.8 $\pm$ 67.6 & 88.9 $\pm$ 1.5 & 13.3 $\pm$ 3.8 & 61.1 $\pm$ 5.8 & -0.46 $\pm$ 0.17 & 2.42 $\pm$ 0.46 & 10.05 $\pm$ 3.87 \\
2009-04-10 18:42:45 & 4.9 $\pm$ 1.4 & 107.9 $\pm$ 4.4 & 4.4 $\pm$ 2.4 & 45.5 $\pm$ 3.4 & -1.91 $\pm$ 1.62 & 1.33 $\pm$ 0.35 & 6.52 $\pm$ 1.35 \\
\enddata
\end{deluxetable*}

Some of the events studied in this paper with likely associations have been recorded and analyzed also from the ground. Examples are the famous Chelyabinsk fireball (2013-02-15 03:20:33) or the event that produced the Flensburg meteorite (2019-09-12 12:49:48), whose orbits were studied in detail and can be computed accurately assuming much smaller uncertainties than those in this work. Other events such as 2020-11-28 16:34:11 were recorded and analyzed by the Japanese SonotaCo consortium system \citep{sonotaco2009meteor}, obtaining results that are compatible with our proposed association with Andromedids meteor shower, although it does not meet the criterion of orbit dissimilarity over time \citep{sonotaco2020fireball}.

The atmospheric entry of asteroid 2008 TC3 was recorded by USG sensors as well (2008-10-07 02:45:40), presenting values contrary to the ones obtained using multiple ground-based observations \citep{Farnocchia2017Icar..294..218F}, as pointed out by \citet{Devillepoix2019MNRAS.483.5166D}. However, we found that this error may simply be a typo in transcribing the values to the CNEOS website. \citet{Farnocchia2017Icar..294..218F} states that the atmospheric flight of the asteroid seemed to come from the north (negative velocity in the z-direction). This does not agree with the z component of the velocity vector appearing in CNEOS $[-9, 9, 3.8]$ km/s, which is positive. By doing the calculations with -3.8 km/s, the results then match reasonably well with \citet{Farnocchia2017Icar..294..218F}, as seen in Table \ref{tab:miscel}.

More recently, on March 11, 2022, asteroid 2022 EB5 was observed prior to its impact with the atmosphere and also recorded by USG sensors. From the data published on the CNEOS web portal, we calculated the orbital elements with the same methodology applied to this work. Table \ref{tab:miscel} shows that the results are in relative good agreement with those reported by the Minor Planet Center (MPC). The largest difference is found in the estimation of the semi-major axis, even though the value is within the uncertainty margins. Nevertheless, the association appears to be a good candidate, except that it only meets a dissimilarity criterion for 3,400 years, as shown in Table \ref{tab:NEA}. Similarly occurs for asteroid 2018 LA, however, the dynamic relation was discarded for not exceeding the expected randomly. These associations could have been reliably established if a distribution of clones within the uncertainties were propagated backward in time, instead of just the median value. We have not performed this fine calculation for all events because our intention was to calculate a reliable minimum percentage of associations with near-Earth objects. However, for the particular case of 2022 EB5, we computed the following minimum orbit dissimilarity values held below the threshold over 10,000 years: $D_{SH}^{min}=0.005$, $D_{J}^{min}=0.033$, $D_{D}^{min}=0.002$ and $D_{H}^{min}=0.005$. For the case of the asteroid 2019 MO impact, the values provided by CNEOS do not allow to establish a clear association, as can be seen in the difference of the $\Omega$ and $\omega$ in Table \ref{tab:miscel}.

\begin{deluxetable*}{lccccccc}
\tablecaption{Semi-major axis, eccentricity, inclination, argument of perihelion and ascending node for the 2008-10-07 02:45:40 and 2022-03-11 21:22:46 produced by the impact of asteroid 2022 EB5, 2019 MO, 2018 LA and 2008 TC3. FAR = \citet{Farnocchia2017Icar..294..218F}. JEN = \citet{jenniskens2021impact}. COR$^*$ = corrected CNEOS $V_z$ sign. \label{tab:miscel}}
\tablewidth{0pt}
\tablehead{
\colhead{Event} & NEA & \colhead{Data} & \colhead{$a\, (AU)$}  & \colhead{$e$} &
\colhead{$i\,(^{\circ})$} & \colhead{$\omega\,(^{\circ})$} & \colhead{$\Omega\,(^{\circ})$} 
}
\startdata
    2022-03-11 21:22:46 & 2022 EB5 & MPC     & 2.8194097       & 0.6851332       & 10.40900       & 222.40750      & 350.99733      \\
     & & CNEOS   & 2.20$\pm$0.82      & 0.59$\pm$0.16     & 9.26$\pm$3.34     & 222.14$\pm$6.21  & 350.976$\pm$0.001        \\
    2019-06-22 21:25:48 & 2019 MO & MPC     & 2.4582908       & 0.6181381       & 1.54135      & 216.73545      & 91.04007      \\
     & & CNEOS   & 2.45$\pm$1.22      & 0.60$\pm$0.22     & 0.70$\pm$0.61     & 29.57$\pm$16.71  & 270.913$\pm$0.001        \\
    2018-06-02 16:44:12 & 2018 LA & JEN     & 1.37640       & 0.431861       & 4.29741       & 256.04869      & 71.869605      \\
     & & CNEOS   & 1.28$\pm$0.30      & 0.40$\pm$0.15     & 4.44$\pm$1.53     & 261.21$\pm$11.36  & 71.850$\pm$0.001        \\   
    2008-10-07 02:45:40 & 2008 TC3 & FAR & 1.3082050 & 0.3120674 & 2.542215 & 234.448925 & 194.1011436  \\
     & & CNEOS & 1.59$\pm$0.23 &	0.42$\pm$0.09 & 4.09$\pm$1.83 & 39.61$\pm$13.58 &	14.089$\pm$0.001 \\
     & & COR$^*$ & 1.30$\pm$0.14 &	0.35$\pm$0.08 & 2.64$\pm$1.04 & 241.84$\pm$17.03 & 194.089 $\pm$ 0.001 \\
\enddata
\end{deluxetable*}

This kind of mismatch and the lack of uncertainty makes the USG sensor database not very reliable at first glance. However, with the approach we propose in this paper, we can at least narrow down the probabilities of a large fireball to be associated with a meteoroid stream or a NEO.

\section{Conclusions}
We examined the CNEOS database superbolide events recorded by US Government sensors with sufficient information to calculate their orbits. However, since there is a lack of uncertainty values in the reported database, we assumed a deviation in the provided parameters based on previous studies that have compared some of these detections with ground-based observations. We developed a new statistical procedure to overtake this challenge. We generated thousands of clones within these error margins and analyzed the associations of each with meteoroid streams and NEOs, thus obtaining a probabilistic estimate of their origins. Following that approach, we have reached the following conclusions:
\begin{itemize}

    \item Among the 887 events in the CNEOS fireball database, only 255 contain enough information to allow the calculation of the heliocentric orbit, so we concentrated our effort in those specific events. 
    
    \item Using a common methodology, performing backwards integration of the orbits over 10,000 years, applying four dissimilarity criteria and tracking their evolution over time, and estimating a false positive rate, we find that 58 of these events probably have an asteroidal or cometary origin. This number corresponds to about 23\% of the total number of meter-sized projectiles producing large bolides. This suggests that roughly one of each four superbolides are originated by near-Earth objects.
    
    \item Based on the height and velocity of each event provided by CNEOS, we estimated the tensile strength for classifying the projectiles according to their bulk density. We found that 61.9\% is rocky or rocky-iron, 35.7\% is carbonaceous and 2.4\% is cometary. Regarding their heliocentric orbits, 85.5\% present orbital elements typical of asteroids while 9.8\% belong to the JFCs.

    \item We identified 5 events with hyperbolic orbits, 3 of them keeping this condition in the uncertainty range and showing high tensile strength values. We confirm the orbital elements of the announced as the first interstellar superbolide (2014-01-08 17:05:34). Another possible events with unbound heliocentric orbits are 2017-03-09 04:16:37 and 2009-04-10 18:42:45. Corroborating these results would imply that at least about 1\% of the large meteoroids impacting our atmosphere could be interstellar in origin, which may be biased in size by the harsh interstellar medium.
    
    \item We find especially reliable the $D_{J}$ and $D_D$ criteria, since having a different theoretical approach they results in similar performances. We also check the robustness of the threshold values for each criterion in relation to variation of the number of associations.
    
    \item Given the typical decoherence time scales for meteoroid streams and how fast large meteoroids can segregate from their respective streams or parent bodies, the mere existence of meter-sized projectiles from NEOs points out physical processes capable of producing a transient population of large meteoroids in relatively short timescales to avoid orbital decoherence. We envision that the disruption of crumble asteroids and comets, often due to impacts or tidal forces during close approaches could be the underlying reason. 
    
    \item The scientific interest in identifying such sources of large bolides is enormous, given that most of these events are potentially delivering meteorites to the Earth and the Moon, so they give the chance to establish direct links between asteroids, meteoroid streams and meteorites. We expect that the new capabilities to identify the sources of meteorite-dropping bolides serve as get new free-delivered samples from hazardous asteroids and comets crossing the near-Earth region. Scientific opportunity is enormous as the bulk elemental composition, mineralogy and physical properties of these projectiles can be inferred from these new meteorites.
    
    \item In addition, our capacity to identify dangerous meteoroid streams, capable to generate meter-sized projectiles can be useful to evaluate the specific risk for extravehicular operations in future Lunar exploration, like the Artemis missions \citep{Artemieva2008SoSyR..42..329A, Moorhead2020JSpRo..57..160M}. Similar goal can be achieved from the study of large bolides on Mars for the future Mars Sample Return initiative, or other manned missions.
    
    \item Our findings imply that both NEOs and certain meteoroid streams are potential producers of meter-sized hazardous projectiles. Identifying the main NEOs producing large bolides might allow us to study possible encounter scenarios and develop palliative strategies and ways to sample these materials for further study \citep{Abell2021BAAS...53d.270A, Barbee2021BAAS...53d.117B, ElisaEloy2022AcAau.192..402S}.  

\end{itemize}

Finally, we also emphasize the great contribution that would represent for the scientific planetary protection community if the events published by CNEOS were accompanied by uncertainties or some measures of the quality of the detection. Satellite records may play a crucial role in the study of Earth colliding meter-sized rocks, a consequence of the continuous decay of asteroids and comets, both to prepare campaigns to search for possible dropped meteorites and to better understand the physical processes delivering space objects to our planet.

\section*{Acknowledgements}
This project has received funding from the European Research Council (ERC) under the European Union’s Horizon 2020 research and innovation programme (grant agreement No. 865657) for the project “Quantum Chemistry on Interstellar Grains” (QUANTUMGRAIN). A total of 2,500 hours of supercomputing time has been used to perform the orbital studies and backwards integrations using CSUC facilities. JMT-R, EPA and AR acknowledge financial support from the FEDER/Ministerio de Ciencia e Innovación – Agencia Estatal de Investigación (PGC2018-097374-B-I00, PI: JMT-R; CTQ2017-89132-P, PI: AR). AR is indebted to the “Ramón y Cajal” program and DIUE (project 2017SGR1323). EPA thanks Carlos Gascón for helpful comments.

\section*{Availability of data}
The data underlying this article will be shared on reasonable request to the corresponding author.

\section{Appendix information}
\startlongtable
\begin{deluxetable*}{llccccccccccccc}
\tabletypesize{\tiny}
\tablecaption{All events with association and the time evolution of the $D_{SH}$ and $D_{D}$ criterion over 10,000 years backwards. The value shown is the mean of each 1,000-year step. \label{tab:appendix}}
\tablewidth{0pt}
\tablehead{
\colhead{Event}               & \colhead{Association}             & \colhead{Crit.}    & \colhead{0 yr}     & \colhead{-1 kyr} & \colhead{-2 kyr} & \colhead{-3 kyr} & \colhead{-4 kyr} & \colhead{-5 kyr} & \colhead{-6 kyr} & \colhead{-7 kyr} & \colhead{-8 kyr} &\colhead{ -9 kyr} & \colhead{-10 kyr}
}
\startdata
2022-02-17 03:53:24 & gamma Delphinids & $D_{SH}$ & 0.25 & 0.21 & 0.29 & 0.29 & 0.28 & 0.36 & 0.46 & 0.55 & 0.61 & 0.66 & 0.86 &  \\
 &  & $D_{D}$ & 0.13 & 0.09 & 0.11 & 0.12 & 0.11 & 0.13 & 0.15 & 0.18 & 0.2 & 0.22 & 0.31 &  \\
2021-11-28 18:06:50 & 2016 WN8 & $D_{SH}$ & 0.24 & 0.24 & 0.23 & 0.23 & 0.22 & 0.22 & 0.23 & 0.23 & 0.24 & 0.25 & 0.26 &  \\
 &  & $D_{D}$ & 0.14 & 0.15 & 0.15 & 0.15 & 0.15 & 0.15 & 0.15 & 0.14 & 0.14 & 0.14 & 0.15 &  \\
2021-11-17 15:53:21 & 2012 KN11 & $D_{SH}$ & 0.11 & 0.12 & 0.15 & 0.19 & 0.21 & 0.21 & 0.19 & 0.19 & 0.21 & 0.27 & 0.33 &  \\
 &  & $D_{D}$ & 0.07 & 0.07 & 0.07 & 0.08 & 0.08 & 0.08 & 0.08 & 0.08 & 0.09 & 0.1 & 0.12 &  \\
2021-04-02 15:52:58 & gamma Taurids & $D_{SH}$ & 0.28 & 0.29 & 0.34 & 0.4 & 0.43 & 0.43 & 0.39 & 0.32 & 0.26 & 0.23 & 0.26 &  \\
 &  & $D_{D}$ & 0.12 & 0.11 & 0.13 & 0.14 & 0.14 & 0.14 & 0.13 & 0.11 & 0.1 & 0.1 & 0.12 &  \\
2021-03-05 13:50:01 & 2005 CV69 & $D_{SH}$ & 0.14 & 0.14 & 0.14 & 0.15 & 0.18 & 0.22 & 0.27 & 0.3 & 0.31 & 0.31 & 0.29 &  \\
 &  & $D_{D}$ & 0.07 & 0.07 & 0.06 & 0.06 & 0.07 & 0.09 & 0.11 & 0.13 & 0.13 & 0.11 & 0.09 &  \\
2020-12-29 20:32:22 & Andromedids & $D_{SH}$ & 0.21 & 0.2 & 0.2 & 0.19 & 0.2 & 0.21 & 0.23 & 0.25 & 0.27 & 0.29 & 0.31 &  \\
 &  & $D_{D}$ & 0.18 & 0.18 & 0.17 & 0.17 & 0.17 & 0.17 & 0.17 & 0.17 & 0.18 & 0.18 & 0.18 &  \\
2020-10-26 15:09:10 & gamma Triangulids & $D_{SH}$ & 0.29 & 0.29 & 0.27 & 0.25 & 0.23 & 0.21 & 0.18 & 0.15 & 0.13 & 0.12 & 0.14 &  \\
 &  & $D_{D}$ & 0.14 & 0.14 & 0.14 & 0.16 & 0.18 & 0.21 & 0.23 & 0.24 & 0.24 & 0.23 & 0.21 &  \\
2020-07-12 07:50:32 & Daytime theta Aurigids & $D_{SH}$ & 0.16 & 0.24 & 0.31 & 0.54 & 0.58 & 0.61 & 0.46 & 0.27 & 0.26 & 0.34 & 0.46 &  \\
 &  & $D_{D}$ & 0.07 & 0.09 & 0.13 & 0.21 & 0.22 & 0.24 & 0.23 & 0.12 & 0.1 & 0.12 & 0.17 &  \\
 & 2021 AU4 & $D_{SH}$ & 0.15 & 0.15 & 0.16 & 0.16 & 0.17 & 0.17 & 0.18 & 0.18 & 0.19 & 0.19 & 0.2 &  \\
 &  & $D_{D}$ & 0.08 & 0.08 & 0.09 & 0.1 & 0.1 & 0.11 & 0.11 & 0.11 & 0.12 & 0.12 & 0.12 &  \\
2020-03-04 20:25:59 & C/1905 F1 & $D_{SH}$ & 0.24 & 0.31 & 0.41 & 0.4 & 0.29 & 0.21 & 0.25 & 0.3 & 0.29 & 0.32 & 0.42 &  \\
 &  & $D_{D}$ & 0.1 & 0.11 & 0.14 & 0.13 & 0.1 & 0.08 & 0.09 & 0.1 & 0.1 & 0.11 & 0.14 &  \\
 & 2020 DF4 & $D_{SH}$ & 0.11 & 0.1 & 0.1 & 0.1 & 0.1 & 0.1 & 0.1 & 0.09 & 0.09 & 0.09 & 0.09 &  \\
 &  & $D_{D}$ & 0.04 & 0.05 & 0.05 & 0.05 & 0.05 & 0.05 & 0.05 & 0.05 & 0.05 & 0.04 & 0.04 &  \\
2019-09-14 12:39:34 & alpha Geminids & $D_{SH}$ & 0.25 & 0.23 & 0.23 & 0.23 & 0.23 & 0.24 & 0.24 & 0.25 & 0.25 & 0.24 & 0.24 &  \\
 &  & $D_{D}$ & 0.22 & 0.21 & 0.21 & 0.21 & 0.21 & 0.21 & 0.21 & 0.21 & 0.21 & 0.2 & 0.2 &  \\
2019-09-12 12:49:48 & Corvids & $D_{SH}$ & 0.26 & 0.24 & 0.24 & 0.25 & 0.26 & 0.27 & 0.27 & 0.25 & 0.23 & 0.22 & 0.22 &  \\
 &  & $D_{D}$ & 0.12 & 0.11 & 0.11 & 0.11 & 0.11 & 0.11 & 0.11 & 0.11 & 0.1 & 0.1 & 0.1 &  \\
 & 300P/Catalina & $D_{SH}$ & 0.13 & 0.12 & 0.12 & 0.12 & 0.12 & 0.13 & 0.12 & 0.12 & 0.12 & 0.12 & 0.12 &  \\
 &  & $D_{D}$ & 0.16 & 0.15 & 0.16 & 0.16 & 0.16 & 0.17 & 0.17 & 0.16 & 0.16 & 0.16 & 0.16 &  \\
2019-07-23 20:42:58 & Corvids & $D_{SH}$ & 0.17 & 0.18 & 0.19 & 0.11 & 0.07 & 0.09 & 0.15 & 0.2 & 0.23 & 0.25 & 0.28 &  \\
 &  & $D_{D}$ & 0.06 & 0.07 & 0.08 & 0.04 & 0.03 & 0.03 & 0.05 & 0.07 & 0.08 & 0.08 & 0.09 &  \\
2019-06-22 21:25:48 & Corvids & $D_{SH}$ & 0.22 & 0.22 & 0.22 & 0.21 & 0.19 & 0.18 & 0.17 & 0.17 & 0.17 & 0.18 & 0.21 &  \\
 &  & $D_{D}$ & 0.08 & 0.08 & 0.08 & 0.08 & 0.08 & 0.07 & 0.07 & 0.07 & 0.07 & 0.08 & 0.09 &  \\
2019-05-19 14:47:03 & Corvids & $D_{SH}$ & 0.19 & 0.19 & 0.21 & 0.22 & 0.24 & 0.26 & 0.28 & 0.3 & 0.3 & 0.3 & 0.3 &  \\
 &  & $D_{D}$ & 0.08 & 0.08 & 0.08 & 0.08 & 0.09 & 0.09 & 0.1 & 0.11 & 0.1 & 0.1 & 0.1 &  \\
2019-03-19 02:06:39 & C/1905 F1 & $D_{SH}$ & 0.38 & 0.42 & 0.47 & 0.42 & 0.4 & 0.46 & 0.49 & 0.48 & 0.53 & 0.54 & 0.55 &  \\
 &  & $D_{D}$ & 0.13 & 0.14 & 0.15 & 0.14 & 0.13 & 0.15 & 0.16 & 0.17 & 0.18 & 0.18 & 0.19 &  \\
2019-02-01 18:17:10 & 2015 AO43 & $D_{SH}$ & 0.13 & 0.13 & 0.13 & 0.14 & 0.14 & 0.15 & 0.15 & 0.15 & 0.16 & 0.16 & 0.17 &  \\
 &  & $D_{D}$ & 0.08 & 0.07 & 0.08 & 0.08 & 0.08 & 0.08 & 0.09 & 0.09 & 0.09 & 0.09 & 0.09 &  \\
2018-10-05 00:27:04 & gamma Taurids & $D_{SH}$ & 0.1 & 0.11 & 0.12 & 0.14 & 0.16 & 0.17 & 0.16 & 0.14 & 0.11 & 0.11 & 0.13 &  \\
 &  & $D_{D}$ & 0.08 & 0.09 & 0.09 & 0.1 & 0.1 & 0.11 & 0.1 & 0.09 & 0.08 & 0.07 & 0.09 &  \\
2018-09-25 14:10:33 & Corvids & $D_{SH}$ & 0.24 & 0.22 & 0.25 & 0.26 & 0.25 & 0.24 & 0.21 & 0.21 & 0.27 & 0.36 & 0.42 &  \\
 &  & $D_{D}$ & 0.16 & 0.13 & 0.14 & 0.15 & 0.15 & 0.15 & 0.14 & 0.13 & 0.14 & 0.16 & 0.18 &  \\
2018-06-26 17:51:53 & Corvids & $D_{SH}$ & 0.25 & 0.26 & 0.28 & 0.3 & 0.33 & 0.36 & 0.38 & 0.42 & 0.45 & 0.48 & 0.52 &  \\
 &  & $D_{D}$ & 0.16 & 0.16 & 0.16 & 0.16 & 0.17 & 0.17 & 0.18 & 0.19 & 0.2 & 0.21 & 0.21 &  \\
2018-01-15 02:18:38 & P/2003 T12 & $D_{SH}$ & 0.33 & 0.33 & 0.24 & 0.21 & 0.33 & 0.27 & 0.34 & 0.21 & 0.26 & 0.19 & 0.15 &  \\
 &  & $D_{D}$ & 0.14 & 0.14 & 0.1 & 0.08 & 0.11 & 0.13 & 0.14 & 0.1 & 0.12 & 0.12 & 0.1 &  \\
2017-12-15 13:14:37 & 2008 ON13 & $D_{SH}$ & 0.2 & 0.2 & 0.2 & 0.21 & 0.21 & 0.21 & 0.21 & 0.21 & 0.2 & 0.2 & 0.2 &  \\
 &  & $D_{D}$ & 0.09 & 0.08 & 0.08 & 0.08 & 0.08 & 0.08 & 0.08 & 0.08 & 0.09 & 0.09 & 0.09 &  \\
 & Northern chi Orionids & $D_{SH}$ & 0.2 & 0.21 & 0.2 & 0.2 & 0.2 & 0.19 & 0.19 & 0.19 & 0.19 & 0.2 & 0.2 &  \\
 &  & $D_{D}$ & 0.13 & 0.15 & 0.14 & 0.14 & 0.14 & 0.14 & 0.14 & 0.14 & 0.14 & 0.14 & 0.14 &  \\
2017-07-13 09:30:36 & gamma Taurids & $D_{SH}$ & 0.18 & 0.15 & 0.13 & 0.13 & 0.14 & 0.15 & 0.15 & 0.15 & 0.16 & 0.15 & 0.15 &  \\
 &  & $D_{D}$ & 0.07 & 0.06 & 0.06 & 0.06 & 0.06 & 0.06 & 0.06 & 0.06 & 0.06 & 0.06 & 0.06 &  \\
2017-06-23 20:21:55 & Southern October delta Arietids & $D_{SH}$ & 0.18 & 0.17 & 0.18 & 0.18 & 0.18 & 0.18 & 0.18 & 0.18 & 0.19 & 0.19 & 0.19 &  \\
 &  & $D_{D}$ & 0.15 & 0.14 & 0.15 & 0.14 & 0.14 & 0.14 & 0.14 & 0.15 & 0.15 & 0.15 & 0.14 &  \\
2017-05-24 07:03:03 & gamma Taurids & $D_{SH}$ & 0.15 & 0.14 & 0.14 & 0.14 & 0.14 & 0.14 & 0.15 & 0.15 & 0.15 & 0.16 & 0.16 &  \\
 &  & $D_{D}$ & 0.1 & 0.1 & 0.1 & 0.1 & 0.1 & 0.1 & 0.1 & 0.1 & 0.1 & 0.1 & 0.1 &  \\
2017-02-18 19:48:29 & alpha Virginids & $D_{SH}$ & 0.24 & 0.19 & 0.17 & 0.16 & 0.15 & 0.15 & 0.15 & 0.16 & 0.17 & 0.18 & 0.19 &  \\
 &  & $D_{D}$ & 0.1 & 0.08 & 0.07 & 0.07 & 0.07 & 0.07 & 0.07 & 0.07 & 0.07 & 0.07 & 0.07 &  \\
2015-07-19 07:06:26 & 300P/Catalina & $D_{SH}$ & 0.3 & 0.29 & 0.24 & 0.21 & 0.29 & 0.34 & 0.31 & 0.21 & 0.17 & 0.27 & 0.33 &  \\
 &  & $D_{D}$ & 0.1 & 0.1 & 0.09 & 0.08 & 0.1 & 0.11 & 0.1 & 0.08 & 0.07 & 0.11 & 0.12 &  \\
2015-04-21 01:42:51 & h Virginids & $D_{SH}$ & 0.29 & 0.27 & 0.24 & 0.21 & 0.19 & 0.17 & 0.17 & 0.17 & 0.18 & 0.19 & 0.21 &  \\
 &  & $D_{D}$ & 0.14 & 0.14 & 0.13 & 0.12 & 0.12 & 0.12 & 0.11 & 0.11 & 0.12 & 0.12 & 0.13 &  \\
2015-04-08 04:06:31 & gamma Triangulids & $D_{SH}$ & 0.33 & 0.32 & 0.32 & 0.29 & 0.24 & 0.19 & 0.16 & 0.17 & 0.18 & 0.19 & 0.2 &  \\
 &  & $D_{D}$ & 0.18 & 0.16 & 0.16 & 0.15 & 0.14 & 0.13 & 0.1 & 0.1 & 0.1 & 0.1 & 0.1 &  \\
 & 182P/LONEOS & $D_{SH}$ & 0.34 & 0.41 & 0.52 & 0.57 & 0.53 & 0.46 & 0.51 & 0.68 & 0.85 & 0.97 & 1.0 &  \\
 &  & $D_{D}$ & 0.14 & 0.17 & 0.22 & 0.24 & 0.22 & 0.19 & 0.2 & 0.25 & 0.3 & 0.34 & 0.35 &  \\
2014-11-26 23:16:51 & 1996 JG & $D_{SH}$ & 0.1 & 0.12 & 0.12 & 0.12 & 0.12 & 0.12 & 0.12 & 0.12 & 0.13 & 0.13 & 0.14 &  \\
 &  & $D_{D}$ & 0.06 & 0.07 & 0.07 & 0.07 & 0.07 & 0.07 & 0.07 & 0.08 & 0.08 & 0.08 & 0.08 &  \\
 & Daytime delta Scorpiids & $D_{SH}$ & 0.26 & 0.25 & 0.24 & 0.24 & 0.24 & 0.24 & 0.23 & 0.22 & 0.22 & 0.22 & 0.23 &  \\
 &  & $D_{D}$ & 0.17 & 0.16 & 0.16 & 0.15 & 0.15 & 0.15 & 0.15 & 0.14 & 0.14 & 0.15 & 0.15 &  \\
2014-11-04 20:13:30 & gamma Taurids & $D_{SH}$ & 0.1 & 0.1 & 0.09 & 0.09 & 0.1 & 0.09 & 0.1 & 0.1 & 0.11 & 0.11 & 0.11 &  \\
 &  & $D_{D}$ & 0.09 & 0.08 & 0.08 & 0.09 & 0.09 & 0.09 & 0.09 & 0.09 & 0.09 & 0.09 & 0.09 &  \\
2014-06-28 02:40:07 & tau Herculids & $D_{SH}$ & 0.23 & 0.2 & 0.19 & 0.2 & 0.19 & 0.17 & 0.17 & 0.17 & 0.19 & 0.21 & 0.24 &  \\
 &  & $D_{D}$ & 0.16 & 0.15 & 0.14 & 0.15 & 0.14 & 0.13 & 0.12 & 0.12 & 0.12 & 0.12 & 0.13 &  \\
2014-05-16 20:06:28 & beta Cancrids & $D_{SH}$ & 0.22 & 0.22 & 0.21 & 0.21 & 0.21 & 0.19 & 0.16 & 0.14 & 0.12 & 0.11 & 0.12 &  \\
 &  & $D_{D}$ & 0.08 & 0.09 & 0.08 & 0.08 & 0.09 & 0.1 & 0.11 & 0.13 & 0.17 & 0.23 & 0.3 &  \\
2013-12-23 08:30:57 & alpha Geminids & $D_{SH}$ & 0.23 & 0.23 & 0.23 & 0.23 & 0.22 & 0.22 & 0.22 & 0.22 & 0.22 & 0.22 & 0.22 &  \\
 &  & $D_{D}$ & 0.14 & 0.15 & 0.15 & 0.15 & 0.15 & 0.15 & 0.15 & 0.15 & 0.15 & 0.15 & 0.14 &  \\
2013-12-08 03:10:09 & October gamma Cetids & $D_{SH}$ & 0.18 & 0.19 & 0.21 & 0.24 & 0.23 & 0.22 & 0.2 & 0.19 & 0.17 & 0.16 & 0.15 &  \\
 &  & $D_{D}$ & 0.07 & 0.09 & 0.11 & 0.13 & 0.12 & 0.11 & 0.09 & 0.08 & 0.07 & 0.06 & 0.06 &  \\
2013-04-30 08:40:38 & October gamma Cetids & $D_{SH}$ & 0.26 & 0.26 & 0.24 & 0.23 & 0.22 & 0.2 & 0.2 & 0.19 & 0.18 & 0.18 & 0.19 &  \\
 &  & $D_{D}$ & 0.15 & 0.14 & 0.13 & 0.12 & 0.12 & 0.1 & 0.09 & 0.07 & 0.06 & 0.06 & 0.07 &  \\
2013-04-21 06:23:12 & Corvids & $D_{SH}$ & 0.25 & 0.28 & 0.33 & 0.33 & 0.3 & 0.28 & 0.27 & 0.25 & 0.24 & 0.24 & 0.25 &  \\
 &  & $D_{D}$ & 0.19 & 0.19 & 0.2 & 0.2 & 0.21 & 0.21 & 0.21 & 0.21 & 0.22 & 0.22 & 0.23 &  \\
2012-07-25 07:48:20 & 2018 BJ5 & $D_{SH}$ & 0.15 & 0.17 & 0.19 & 0.19 & 0.19 & 0.19 & 0.18 & 0.19 & 0.2 & 0.2 & 0.2 &  \\
 &  & $D_{D}$ & 0.22 & 0.3 & 0.35 & 0.33 & 0.33 & 0.31 & 0.29 & 0.28 & 0.28 & 0.29 & 0.28 &  \\
2012-05-15 11:04:17 & 2020 KF6 & $D_{SH}$ & 0.13 & 0.13 & 0.13 & 0.13 & 0.13 & 0.13 & 0.13 & 0.12 & 0.12 & 0.12 & 0.13 &  \\
 &  & $D_{D}$ & 0.34 & 0.37 & 0.35 & 0.33 & 0.31 & 0.29 & 0.26 & 0.17 & 0.12 & 0.2 & 0.39 &  \\
2010-12-25 23:24:00 & 2014 XO7 & $D_{SH}$ & 0.21 & 0.52 & 0.69 & 0.29 & 0.48 & 0.27 & 0.47 & 0.38 & 0.52 & 0.77 & 0.38 &  \\
 &  & $D_{D}$ & 0.12 & 0.19 & 0.24 & 0.13 & 0.17 & 0.13 & 0.16 & 0.15 & 0.21 & 0.26 & 0.19 &  \\
2010-02-28 22:24:50 & 79P/du Toit-Hartley & $D_{SH}$ & 0.25 & 0.25 & 0.25 & 0.25 & 0.25 & 0.26 & 0.26 & 0.26 & 0.26 & 0.26 & 0.26 &  \\
 &  & $D_{D}$ & 0.15 & 0.15 & 0.15 & 0.15 & 0.15 & 0.15 & 0.15 & 0.15 & 0.14 & 0.14 & 0.13 &  \\
2008-12-24 15:51:58 & P/2003 T12 & $D_{SH}$ & 0.32 & 0.43 & 0.61 & 0.58 & 0.39 & 0.41 & 0.6 & 0.64 & 0.5 & 0.4 & 0.56 &  \\
 &  & $D_{D}$ & 0.12 & 0.14 & 0.2 & 0.19 & 0.14 & 0.14 & 0.19 & 0.2 & 0.17 & 0.14 & 0.19 &  \\
2008-11-21 00:26:44 & pi Leonids & $D_{SH}$ & 0.25 & 0.23 & 0.2 & 0.19 & 0.19 & 0.23 & 0.26 & 0.3 & 0.33 & 0.34 & 0.34 &  \\
 &  & $D_{D}$ & 0.09 & 0.08 & 0.07 & 0.07 & 0.08 & 0.09 & 0.1 & 0.12 & 0.12 & 0.12 & 0.11 &  \\
2008-11-18 09:41:51 & iota Cygnids & $D_{SH}$ & 2.01 & 2.01 & 2.0 & 1.99 & 1.98 & 1.97 & 1.97 & 1.96 & 1.95 & 1.94 & 1.93 &  \\
 &  & $D_{D}$ & 0.92 & 0.92 & 0.91 & 0.9 & 0.89 & 0.88 & 0.87 & 0.86 & 0.85 & 0.85 & 0.84 &  \\
2008-06-27 02:01:23 & Daytime delta Scorpiids & $D_{SH}$ & 0.33 & 0.33 & 0.33 & 0.31 & 0.28 & 0.24 & 0.2 & 0.2 & 0.22 & 0.26 & 0.29 &  \\
 &  & $D_{D}$ & 0.17 & 0.16 & 0.16 & 0.15 & 0.15 & 0.14 & 0.13 & 0.13 & 0.13 & 0.14 & 0.14 &  \\
2007-09-22 17:57:12 & Southern omega Scorpiids & $D_{SH}$ & 0.13 & 0.35 & 0.4 & 0.4 & 0.77 & 0.41 & 0.44 & 0.45 & 0.27 & 0.42 & 0.33 &  \\
 &  & $D_{D}$ & 0.13 & 0.15 & 0.15 & 0.19 & 0.25 & 0.2 & 0.16 & 0.18 & 0.14 & 0.15 & 0.2 &  \\
2007-06-11 09:47:05 & 34D/Gale & $D_{SH}$ & 0.3 & 0.29 & 0.29 & 0.29 & 0.29 & 0.28 & 0.28 & 0.28 & 0.27 & 0.27 & 0.27 &  \\
 &  & $D_{D}$ & 0.55 & 0.56 & 0.54 & 0.52 & 0.5 & 0.49 & 0.47 & 0.46 & 0.46 & 0.45 & 0.44 &  \\
2006-12-09 06:31:12 & Phoenicids & $D_{SH}$ & 0.22 & 0.25 & 0.4 & 0.42 & 0.31 & 0.31 & 0.45 & 0.52 & 0.48 & 0.46 & 0.55 &  \\
 &  & $D_{D}$ & 0.12 & 0.11 & 0.14 & 0.15 & 0.13 & 0.12 & 0.15 & 0.17 & 0.17 & 0.16 & 0.18 &  \\
2006-10-14 18:10:49 & h Virginids & $D_{SH}$ & 0.39 & 0.34 & 0.29 & 0.34 & 0.3 & 0.21 & 0.25 & 0.29 & 0.19 & 0.15 & 0.18 &  \\
 &  & $D_{D}$ & 0.13 & 0.12 & 0.1 & 0.11 & 0.1 & 0.08 & 0.09 & 0.1 & 0.07 & 0.06 & 0.07 &  \\
2006-09-02 04:26:15 & Corvids & $D_{SH}$ & 0.28 & 0.25 & 0.22 & 0.19 & 0.17 & 0.15 & 0.15 & 0.16 & 0.17 & 0.2 & 0.23 &  \\
 &  & $D_{D}$ & 0.13 & 0.12 & 0.11 & 0.1 & 0.09 & 0.09 & 0.08 & 0.08 & 0.09 & 0.09 & 0.1 &  \\
2006-06-07 00:06:28 & May lambda Draconids & $D_{SH}$ & 0.26 & 0.22 & 0.25 & 0.21 & 0.27 & 0.24 & 0.26 & 0.32 & 0.45 & 0.42 & 0.41 &  \\
 &  & $D_{D}$ & 0.16 & 0.12 & 0.1 & 0.08 & 0.13 & 0.09 & 0.09 & 0.13 & 0.21 & 0.19 & 0.17 &  \\
2006-01-10 23:25:28 & alpha Geminids & $D_{SH}$ & 0.28 & 0.28 & 0.29 & 0.29 & 0.29 & 0.29 & 0.29 & 0.29 & 0.3 & 0.3 & 0.31 &  \\
 &  & $D_{D}$ & 0.16 & 0.15 & 0.15 & 0.15 & 0.15 & 0.15 & 0.15 & 0.15 & 0.15 & 0.15 & 0.16 &  \\
2005-12-03 12:45:49 & alpha Capricornids & $D_{SH}$ & 0.26 & 0.25 & 0.25 & 0.25 & 0.23 & 0.24 & 0.23 & 0.23 & 0.26 & 0.27 & 0.31 &  \\
 &  & $D_{D}$ & 0.17 & 0.16 & 0.16 & 0.16 & 0.15 & 0.15 & 0.15 & 0.15 & 0.15 & 0.16 & 0.16 &  \\
2005-06-03 08:15:41 & Daytime kappa Aquariids & $D_{SH}$ & 0.34 & 0.36 & 0.39 & 0.3 & 0.17 & 0.18 & 0.3 & 0.35 & 0.27 & 0.2 & 0.31 &  \\
 &  & $D_{D}$ & 0.11 & 0.12 & 0.15 & 0.11 & 0.07 & 0.07 & 0.13 & 0.16 & 0.13 & 0.09 & 0.12 &  \\
2005-04-19 07:37:47 & Corvids & $D_{SH}$ & 0.26 & 0.23 & 0.24 & 0.28 & 0.3 & 0.29 & 0.25 & 0.2 & 0.19 & 0.23 & 0.27 &  \\
 &  & $D_{D}$ & 0.15 & 0.12 & 0.1 & 0.1 & 0.11 & 0.1 & 0.1 & 0.09 & 0.08 & 0.09 & 0.1 &  \\
2004-10-07 13:14:43 & Southern omega Scorpiids & $D_{SH}$ & 0.19 & 0.2 & 0.19 & 0.19 & 0.18 & 0.19 & 0.19 & 0.2 & 0.21 & 0.24 & 0.25 &  \\
 &  & $D_{D}$ & 0.19 & 0.21 & 0.2 & 0.19 & 0.18 & 0.19 & 0.2 & 0.2 & 0.21 & 0.23 & 0.23 &  \\
2003-11-10 13:54:06 & 182P/LONEOS & $D_{SH}$ & 0.58 & 0.57 & 0.92 & 1.04 & 0.95 & 0.78 & 0.61 & 0.68 & 0.8 & 0.88 & 0.84 &  \\
 &  & $D_{D}$ & 0.31 & 0.27 & 0.37 & 0.4 & 0.35 & 0.32 & 0.25 & 0.27 & 0.33 & 0.36 & 0.33 &  \\
2003-09-27 12:59:02 & C/1905 F1 & $D_{SH}$ & 0.35 & 0.43 & 0.59 & 0.66 & 0.6 & 0.45 & 0.34 & 0.44 & 0.66 & 0.85 & 0.83 &  \\
 &  & $D_{D}$ & 0.15 & 0.19 & 0.26 & 0.29 & 0.27 & 0.21 & 0.16 & 0.18 & 0.28 & 0.36 & 0.37 &  \\
\enddata
\end{deluxetable*}

\bibliography{sample631}{}
\bibliographystyle{aasjournal}



\end{document}